\documentclass[fleqn,usenatbib]{mnras}
\usepackage{amsmath}
\usepackage{enumitem}
\usepackage{color}
\usepackage{array} 
\usepackage{graphicx}
\usepackage{epstopdf}
\usepackage{xcolor}
\usepackage[labelfont=bf,labelsep=period,justification=raggedright,singlelinecheck=false,font=small,tableposition=top]{caption}
\usepackage{wrapfig}
\usepackage{float}
\usepackage{dblfloatfix}
\usepackage{amssymb}
\usepackage[normalem]{ulem}
\usepackage{hyperref}
\usepackage{undertilde}
\usepackage[T1]{fontenc}
\usepackage{ae,aecompl}
\usepackage[utf8]{inputenc}
\usepackage{comment}
\usepackage[export]{adjustbox}

\def\aap{A\& A}

\def\apjl{ApJL}
\def\apjs{ApJS}

\newcommand{\swift}{{\sc Swift}}
\newcommand{\swifter}{{\sc Sparcs}}
\newcommand{\YKcode}{{\sc HyLight}}
\newcommand{\chimes}{{\sc Chimes}}
\newcommand{\swiftrt}{{\sc Sph-m1rt}}
\newcommand{\radmc}{{\sc Radmc3D}}
\newcommand{\M}{{\sc m1}}
\newcommand{\ifu}{{\sc ifu}}
\newcommand\altaffilmark[1]{$^{#1}$}
\newcommand\altaffiltext[1]{$^{#1}$}
\newcommand{\rt}{{\sc rt}}
\newcommand{\sm}{\small}

\definecolor{orcidlogocol}{HTML}{A6CE39}

\usepackage{scalerel}
\usepackage{tikz}
\usepackage{subfiles}
\usepackage{booktabs}
\usepackage{multirow}
\usetikzlibrary{svg.path}
\usepackage{siunitx}
\tikzset{
  orcidlogo/.pic={
    \fill[orcidlogocol] svg{M256,128c0,70.7-57.3,128-128,128C57.3,256,0,198.7,0,128C0,57.3,57.3,0,128,0C198.7,0,256,57.3,256,128z};
    \fill[white] svg{M86.3,186.2H70.9V79.1h15.4v48.4V186.2z}
                 svg{M108.9,79.1h41.6c39.6,0,57,28.3,57,53.6c0,27.5-21.5,53.6-56.8,53.6h-41.8V79.1z M124.3,172.4h24.5c34.9,0,42.9-26.5,42.9-39.7c0-21.5-13.7-39.7-43.7-39.7h-23.7V172.4z}
                 svg{M88.7,56.8c0,5.5-4.5,10.1-10.1,10.1c-5.6,0-10.1-4.6-10.1-10.1c0-5.6,4.5-10.1,10.1-10.1C84.2,46.7,88.7,51.3,88.7,56.8z};
  }
}

\newcommand\orcidicon[1]{\href{https://orcid.org/#1}{\mbox{\scalerel*{
\begin{tikzpicture}[yscale=-1,transform shape]
\pic{orcidlogo};
\end{tikzpicture}
}{|}}}}

\title[SPARCS]{
 \swifter{}- combining radiation hydrodynamics with non-equilibrium metal chemistry in the \swift{} astrophysical code
}

\author[T. K. Chan et al.]
  {Tsang Keung Chan$^{\orcidicon{0000-0003-2544-054X}}$\altaffilmark{1,2}\thanks{Email: (TKC)tsangkeungchan@cuhk.edu.hk}, Alexander J. Richings$^{\orcidicon{0000-0003-0502-9235}}$\altaffilmark{3,4}, Tom Theuns$^{\orcidicon{0000-0002-3790-9520}}$\altaffilmark{5}, Yuankang Liu$^{\orcidicon{0000-0003-0771-763X}}$\altaffilmark{5},
  \newauthor
  Matthieu Schaller$^{\orcidicon{0000-0002-2395-4902}}$\altaffilmark{6,7}, and
  Mladen Ivkovic\,$^{\orcidicon{0000-0002-3539-3831}}$\altaffilmark{8,9,10}\\
  \altaffiltext{1}{Department of Physics, the Chinese University of Hong Kong, Shatin, Hong Kong, China}\\
\altaffiltext{2}{Department of Astronomy and Astrophysics, the University of Chicago, Chicago, IL60637, USA}\\
  \altaffiltext{3}{Centre for Data Science, Artificial Intelligence and Modelling, University of Hull, Cottingham Road, Hull HU6 7RX, UK}\\
  \altaffiltext{4}{E.A. Milne Centre for Astrophysics, University of Hull, Cottingham Road, Hull HU6 7RX, UK}\\
  \altaffiltext{5}{Institute for Computational Cosmology, Department of Physics, Durham University, South Road, Durham DH1 3LE, UK}\\ 
  \altaffiltext{6}{Lorentz Institute for Theoretical Physics, Leiden University, PO Box 9506, NL-2300 RA Leiden, The Netherlands}\\
  \altaffiltext{7}{Leiden Observatory, Leiden University, PO Box 9513, NL-2300 RA Leiden, The Netherlands}\\
  \altaffiltext{8}{Laboratoire d'astrophysique, \'{E}cole Polytechnique F\'{e}d\'{e}rale de Lausanne (EPFL), 1290 Sauverny, Switzerland}\\
  \altaffiltext{9}{Observatoire de Gen\`{e}ve, Universit\'{e} de Gen\`{e}ve, Chemin Pegasi 51, 1290 Versoix, Switzerland}\\
  \altaffiltext{10}{Department of Computer Science, Durham University, Upper Mountjoy Campus,Stockton Road, Durham, UK}
}

\date{Accepted XXX. Received YYY; in original form ZZZ}

\pubyear{{\the\year{}}}

\begin{document}

\label{firstpage}
\pagerange{\pageref{firstpage}--\pageref{lastpage}}
\maketitle

\begin{abstract}
We present \swifter{}, which combines the moment-based radiative transfer method \swiftrt{} with the non-equilibrium metal chemistry solver \chimes{} in the modern highly-parallel astrophysical code \swift{}. 
\swifter{} enables on-the-fly radiation hydrodynamics simulations,
with multi-frequency ultraviolet radiative transfer coupled with all ionisation states of 11 major elements, in the presence of dust, cosmic ray ionization and heating, and self-gravity. Direct radiation pressure on gas and dust is also accounted for. We validate \swifter{} against analytic solutions and standard photo-ionization codes such as {\sc cloudy} in idealized tests. As an example application, we simulate an ionization front propagating through an inhomogeneous interstellar medium with solar metallicity. We produce mock optical emission line observations with the level population calculation code \YKcode{} and the diagnostic radiative transfer code \radmc{}. We find that non-equilibrium effects and inhomogeneities can boost the low ion fractions by up to an order of magnitude. Possible applications of \swifter{} include studying the dynamical impact of radiation on gas in star-forming regions, and in the interstellar, intergalactic, and circumgalactic medium, as well as interpreting line diagnostics in such environments, and galactic or AGN outflows.

\end{abstract}


\label{firstpage}
\section{Introduction}
\label{sec:introduction}
Radiative processes are essential to the formation of stars and galaxies. While dark matter initiates halo collapse, radiative cooling allows gas to lose pressure support, become self-gravitating, form a disk, and collapse further into dense clouds and stars. This scenario applies equally well to the two-stage theory of galaxy formation of \cite{White78} as to the cold-accretion model of  \cite{Keres05}. The significance of radiative cooling has been well-documented since the early works by \cite{Mestel65} on disk formation, \cite{Field65} on star formation, and \cite{Eggen62} on the collapse of the Milky Way disk.  Radiative processes prevent the formation of galaxies in low mass halos by suppressing gas cooling \citep{Efstathiou92} and photo-heating gas and preventing it from accreting onto or expelling it from such small halos \citep{Benson02, Okamoto08}. Finally, radiative processes also play an important role in setting the efficiency of feedback from stars \citep{Murr10GMCR,Krum13RDust,Hopk20rhd,Rath21SILCC} and black holes \citep{King11AGN,Fauc12AGN} in regulating star-formation in galaxies. 

In addition to being crucial to the dynamics of gas in galaxies, radiative processes provide sensitive diagnostics that allow us to infer the density, temperature, composition (\lq metallicity\rq), and dust content of the gas in addition to properties of the local radiation field. Examples include inferring the star formation rate from the H$\alpha$, [{\rm O}{\sc iii}]$\lambda5007$, or far-infrared flux \cite[e.g.][]{Kenn98SF} of a galaxy or star-forming region, 
using line ratios to determine the density and temperature of the emitting gas \cite[e.g.][]{Kewl19EL}, and characterizing the ambient radiation field with the ``Baldwin, Phillips \& Terlevich'' (BPT) \citep{Baldwin81} and other line-ratio diagrams \cite[e.g.][]{Kewley13}. Furthermore, gas-phase metallicities of galaxies are usually determined with emission line intensities from \ion{H}{II} regions, e.g. \cite{Trem04MassMetal,Pily16abund}.

There are many recent advances of sensitive integral field units ({\sc ifu}), e.g.  {\sc muse} on {\sc vlt} \citep{MUSE}, 
{\sc kcwi} \citep{Morrissey18} on {\sc keck}, {\sc nirspec} \citep{NIRSPEC} on {\sc jwst}, Local Volume Mapper \citep{LVM}, and in the future, {\sc amase} \citep{AMASE}. They allow astronomers to image star-forming regions and galaxies in these characteristic emission lines (see, for example, 
the \lq Pillars of Creation\rq\ imaged with {\sc muse} by \citealt{McLeod15}, or the dense, optically-obscured
regions of star-forming clouds imaged with {\sc nirspec} by \citealt{Jones23}). The {\sc Atacama Large Millimeter Array} ({\sc alma}) complements these optical/near infrared view with radio data, probing the molecular gas associated with these regions \cite[e.g.][]{Wong22}.

Inferring the physical conditions of the star-forming regions from these \ifu{}  or radio data requires detailed modelling of the interaction between radiation and matter.
The {\sc cloudy} spectral synthesis code (\citealt{Ferl17cloudy}, see \citealt{Chatzikos23} for the latest release) sets the benchmark for computing such interaction. {\sc cloudy} performs accurate radiative transfer from a source of radiation to the observer, including absorption, scattering, spontaneous and induced emission by the gas surrounding the source. Alternatively, the {\sc mappings} spectral synthesis code  \citep{Suth18MAPPINGS} also includes the effects of shocks. The {\sc despotic} code can provide fast calculations of temperatures with a one-zone model and an escape probability formalism \citep{Krum14DESPOTIC}. These calculations are restricted to relatively simple geometries of the gas distribution. There are efforts including more realistic 3D geometry, e.g. {\sc mocassin} \citep{Erco03MOCASSIN} and {\sc Messenger Monte Carlo mappings V} \citep{Jin22M3MAPPINGSV}. However, these codes usually assume photo-ionization and thermal steady states.

Observed star-forming regions often have intricate structures that cannot be captured with a simple geometry. These more complicated geometries may affect inferences made from line ratios, for example, when different lines are emitted by spatially distinct regions. Studying this intricate combination of cold, dusty filaments enveloping hot ionized bubbles seen in star-forming regions is interesting in itself, and may reveal how clouds form and eventually disband, for example, due to energetic input (\lq feedback\rq) from stars. To fully exploit all the new data and model how the radiation affects the gas distribution requires radiation hydrodynamical simulations (ideally including magnetic fields and dust physics). Simulations that include hydrodynamics, radiative transfer (\rt), and a detailed chemistry network to account for the interaction of radiation and matter are very computationally demanding.

We briefly summarize how different codes handle radiation hydrodynamics and chemistry. \cite{Gned01OTVET} devised a Variable Eddington Tensor approximation, {\sc otvet}, for radiative transfer, and coupled the radiation with hydrogen and helium ions. Simulating radiative transfer is very expensive, so they also devised a \lq reduced speed of light\rq\ approximation to speed up the calculations. The {\sc traphic} scheme \citep{Pawl08TRAPHIC} discretizes radiation in a finite number of cones and couples the radiation with the hydrogen thermo-chemistry. \cite{Pawl11multifRT} extended the scheme to include hydrogen, helium, and multiple-frequency radiation.

One popular radiative transfer method is the \M\ scheme, a \lq Two-Moment method with Local Eddington Tensor Closure\rq\ \citep{Mine79closure}. The cosmological radiation hydrodynamics codes {\sc ramses} \citep{Teys02RAMSES}, {\sc arepo} \citep{Springel10} and {\sc gizmo} \citep{Hopk15}, all have implementations of the \M\ scheme, as described by \citealt{Rosd13ramsert}, \citealt{Kann19AREPORT} and \citealt{Hopkins19}, respectively. \M\ has the advantage that the \rt\ compute time is independent of the number of sources which is important in situations with many sources, and also allows the calculation to account for the \lq diffuse radiation\rq\ emitted by the gas itself. The reduced speed of light approximation can be applied to significantly accelerate the performance \citep{Gned01OTVET}. Disadvantages of \M{} include diffusivity of the propagated radiation, and the fact that beams of light \lq collide\rq\ when they cross, see e.g. \cite{Rosd13ramsert, Wunsch24,Chan21SPHM1RT} for a discussion of pros and cons of various \rt\ schemes. The \chimes\ photo-chemistry module is also implemented in {\sc arepo} and {\sc gizmo}, but it is not coupled to the \M\ \rt\ scheme. \cite{Katz22RTZ} coupled non-equilibrium thermo-chemistry with the \M\ radiative transfer solver {\sc ramses-rt} \citep{Rosd13ramsert} in {\sc ramses}. Example applications of these codes include the {\sc edge} simulations of the formation of dwarf galaxies in the early universe \citep{Rey25}, which use {\sc ramses}, and the {\sc thesan} simulations of galaxy formation \citep{Kann22THESAN}, which use {\sc arepo-rt} \citep{Kann19AREPORT}.

The combination of \rt\ and a detailed chemistry model is potentially even more relevant for studying the interstellar medium, in particular, the formation of molecular clouds and their interaction with stars that form inside them. \cite{Mell06HIIregion} simulated the dynamics of an \ion{H}{II} region forming in a turbulent molecular cloud using the \lq Conservative, Causal Ray-tracing\rq\ method {\sc C2-ray} \citep{Mell06c2ray}. The {\sc silcc} project uses the {\sc flash} code \citep{Fryxell00} with ray-tracing \rt\ 
\citep{Wunsch21} and the chemistry solver described by \cite{Walch15} and references therein. {\sc torus} \citep{Harr19TORUS} is a  Monte Carlo radiation transfer and hydrodynamics code that can solve photoionization and thermal equilibrium with dust and multiple atomic species (e.g. \citealt{Ali19torus}). {\sc CMacIonize} \citep{VAND18CMacIonize} combined a Monte Carlo photoionisation algorithm with the chemistry of hydrogen, helium, and several coolants. {\sc CMacIonize} was extended to include non-equilibrium thermo-chemistry of H, He, C, N, O, Ne, and S in \cite{McCa24DIGsfgalaxy}.  The {\sc tigress-ncr} project \citep{Kim23TIGRESSNCR} uses the {\sc  Athena MHD} code \citep{Stone08} with adaptive ray-tracing for ultra-violet (UV) photons and the chemistry solver described by \cite{Kim23b}. 
The {\sc starforge} project \citep{Grud21starforge} uses the {\sc gizmo} \citep{Hopk15} code with the {\sc TreeCol} \citep{Clark12} and {\sc lebron} \citep{Hopkins18} \rt\ implementations.

These codes apply different strategies to handle thermochemistry. Some use interpolation tables constructed using {\sc cloudy} and assuming ionization equilibrium, while others solve rate equations and hence account for non-equilibrium effects. Hybrid codes use a combination of these two methods by, for example, solving for the abundances of hydrogen and helium ions in non-equilibrium but assuming that other ions are in ionization equilibrium (e.g. \citealt{Ploe25hybridCHIMES}). Non-equilibrium effects may be important when (1) the cooling time is shorter than the chemical time \citep[e.g.][]{Gnat07NEQ,deAv12NEQ,Rich18AGNH2}, (2) in the presence of strong turbulence or shocks \citep[e.g.][]{Gray17turb,Teci08NEQ}, or (3) when the radiation field varies rapidly in time \citep[e.g.][]{Oppe18AGN}. The non-equilibrium chemistry can also have strong effects on the ionisation states of diffuse ionised gas \citep{McCa24DIGnoeq,McCa24DIGsfgalaxy}.

Non-equilibrium chemistry libraries integrate the rate equations for the ionisation and recombination of ions, and the formation and dissociation of molecules, both in the presence of dust. Examples include the network described by \cite{Glov07chem}, {\sc krome} \citep{Gras14KROME}, and {\sc grackle} \citep{Smith17}, the latter of which is implemented in the public version of \swift{}. In particular, \chimes{} \citep{Rich14CHIMESI,Rich14CHIMESII} tracks the non-equilibrium thermo-chemical evolution of 137 ions and 20 molecules down to temperature $T=10\;{\rm K}$. \chimes{} integrates rate equations to determine the ionization states of ions and the production and destruction rates of molecules, and the associated photo-heating and radiative cooling rates of the gas (accounting for absorption and emission of photons by gas and dust), in the presence of a specified radiation field.

In this paper, we present and test the \swifter{}\footnote{\swifter{} is an acronym for ``{\bf S}moothed {\bf P}article hydrodynamics {\bf A}strophysical method with {\bf R}adiation and non-equilibrium metal {\bf C}hemistry in \swift{}''.} method, which combines a modified \M\ radiative transfer scheme (\swiftrt{}; \citealt{Chan21SPHM1RT}) with the \chimes\ non-equilibrium thermo-chemistry solver \citep{Rich14CHIMESI, Rich14CHIMESII}, implemented in the \swift{} cosmological hydrodynamics code \citep{Scha18SWIFTascl,Scha23SWIFT}. The combination of \swiftrt{} and \chimes{} allows non-equilibrium thermo-chemistry coupling with radiation on the fly. We can also apply the reduced speed of light approximation to accelerate the radiative transfer calculations. In this new implementation, \chimes{} solves self-consistently for the evolution of the radiation field.

As an example application of \swifter{}, we create mock observation images from a simulation snapshot. Hydrogen recombination lines are computed by evaluating the level population of \ion{H}{I} with the \YKcode{} code \citep{Liu25Hylight}. Hydrogen levels are combined with the ionisation states of other ions to create the mock images using \radmc{} \citep{Dullemond12}, a diagnostic Monte Carlo radiative transfer code that accounts for continuum radiative transfer in dusty media, gas line transfer, and gas continuum transfer. Our main motivation for developing \swifter{} is to enable simulations of the interstellar medium ({\sc ism}) of star-forming regions and the circumgalactic medium ({\sc cgm}), incorporating on-the-fly calculations of non-equilibrium thermo-chemistry interactions, along with synthetic observations with detailed emission lines, spectra, and \ifu{} data.

This paper is organised as follows. In \S\ref{sec:methods}, we introduce the concept, physics, and methodology of \swifter{}. In \S\ref{sec:verification}, we validate our multi-frequency radiative transfer with non-equilibrium thermo-chemistry through standard test problems. As an illustration of \swifter{}, we simulate in \S\ref{sec:application} an ionization front propagating through a turbulent interstellar medium with solar metallicity, and demonstrate how the ionization fraction of some ions can be up to an order of magnitude higher due to non-equilibrium effects. In \S\ref{sec:radmc3d}, we apply \YKcode{} and \radmc{} to generate synthetic images of this simulation in different nebular lines. Finally, we assess the parallelization performance of the code in \S\ref{sec:timing}. We summarize and discuss promising future directions for code development, and of applications of \swifter{}, in \S\ref{sec:conclusion}.

\section{Combining  \swiftrt\  with \chimes\ chemistry}
\label{sec:methods}
We begin this section with a brief overview of \chimes\ (\S\ref{sec:CHIMES_method}), and the \swiftrt\ radiative transfer method (\S\ref{sec:SPHM1RT}). In \S\ref{sec:rsl}, we investigate limitations of integrating the propagation of light with a reduced speed to avoid the need for short steps in the time integration. We discuss how we extend and couple \chimes\ with \swiftrt\ in \S\ref{sec:chemistryeq}. Next, we discuss the implementation of dust physics (\S\ref{sec:dustmethod}), radiation pressure (\S\ref{sec:dirradpre}), and radiation injection (\S\ref{sec:inject}). We finish this section with details of the implementation in \swift{} in \S\ref{sec:swiftimplement}.

\subsection{The \chimes{} chemistry module}
\label{sec:CHIMES_method}
{\bf Ch}emistry of the {\bf I}nterstellar {\bf M}edium and {\bf E}xtragalactic {\bf S}ources  (\chimes) is a public\footnote{See Data Availability statement.} non-equilibrium chemistry and photo-heating and radiative cooling module \citep{Rich14CHIMESI, Rich14CHIMESII}. The module
tracks the interaction and time evolution of 157 chemical species, including all ionization levels of eleven elements\footnote{\label{footnote:CHIMESelement}The elements tracked by \chimes{} are H, He, C, N, O, Ne, Mg, Si, S, Ca, Fe. Ions tracked are  all the positive ions of these elements, and in addition ${\rm H^-}$, ${\rm C^-}$, and ${\rm O^-}$.}, as well as some molecules\footnote{The molecules tracked by \chimes{} are ${\rm H_2}$, ${\rm H^+_2}$ , ${\rm H^+_3}$ , ${\rm OH}$, ${\rm H_2O}$, ${\rm C_2}$, ${\rm O_2}$, ${\rm HCO^+}$, ${\rm CH}$, ${\rm CH_2}$, ${\rm CH^+_3}$, ${\rm CO}$, ${\rm CH^+}$, ${\rm CH^+_2}$ ,
${\rm OH^+}$, ${\rm H_2O^+}$, ${\rm H_3O^+}$, ${\rm CO^+}$, ${\rm HOC^+}$, and ${\rm O^+_2}$.}.
The tracked species are selected because they contribute significantly to the cooling/heating of cosmic gas, and because their emission and/or absorption lines are often observed. 

\chimes{} includes rates for the ionization and recombination of these ions, and the formation and dissociation of the molecules in the presence of a specified radiation field and for a specified dust contents. It includes photo-ionization, photo-dissociation, and recombination (see also \S\ref{sec:chemistryeq}), as well as charge-exchange, collisional ionization, cosmic ray ionization and Auger processes. A total of 907 reactions are integrated, as summarized in Table B1 of \cite{Rich14CHIMESI}. However, in this paper we turn off the molecular network in \chimes{} and do not couple the radiative transfer to the molecular network, since we have not yet implemented a prescription for the shielding of Lyman-Werner radiation in the RT module. We intend to improve on this in future.

\chimes{} also accounts for the impact of dust on the chemistry for a given dust content. 
Included dust reactions are ${\rm H}_2$ formation on dust grains, and grain recombination of \ion{H}{II}, \ion{He}{II}, \ion{C}{II}, \ion{O}{II}, \ion{S}{III}, \ion{Fe}{II}, \ion{Mg}{II}, \ion{S}{II}, and \ion{Ca}{II} on dust grains. 

\chimes{} requires the dust temperature as an input parameter as this affects the formation rate of ${\rm H}_2$ on dust grains. However, \cite{Rich14CHIMESI} demonstrated that the dependence is weak in the range $T_{\rm d}=6-50$~K typical of star-forming regions. The rate only depends strongly on dust temperatures outside this range. In this paper, we will be studying the \ion{H}{II} region environments, where the dust temperatures lie within this range (e.g. \citealt{Rela16dust}). We therefore assume a constant dust temperature of 10 K.

We briefly mention important heating and cooling processes included in \chimes{} (see Table 1 of \citealt{Rich14CHIMESI} for the full list of processes).  \chimes{} includes photo heating, H and He excitation, collisional ionization, and recombination cooling. Other processes included in \chimes{} are metal line cooling \citep{Oppe13noneqA}, ${\rm H}_2$ cooling, CO, ${\rm H}_2{\rm O}$, OH cooling \citep{Glov10CO}, grain-surface recombination cooling, photo-electric heating \citep{Bake94photoelectric,Wolf95NAP}, and gas-grain energy exchange. We discuss photo heating and ${\rm H}_2$ photo-electric heating in more detail in \S\ref{sec:chemistryeq}.

\chimes{} integrates the network of 907 interactions between the 157 species together with a thermal energy equation in the presence of a specified radiation field. Additional radiation equations can be added to the system, as described in \S\ref{sec:chemistryeq}.
This demanding integration is performed as follows. \chimes{} first integrates the set of coupled ordinary differential equations (ODEs) using a forward explicit Euler solver without sub-cycling. If the relative change in all variables that are above the absolute tolerance (default value of $1\times 10^{-10}$) is below a certain relative threshold (by default, $10$~per cent), then we keep the explicit solution. Otherwise, \chimes{} integrates the ODEs using the Backward Difference Formula and Newton Iteration scheme {\sc Cvode}, implemented in the Sundials library \citep{hindmarsh2005sundials,gardner2022sundials}\footnote{\url{https://computing.llnl.gov/projects/sundials}}. This allows us to avoid the high computational cost of the {\sc Cvode} solver for particles where there is very little change (for example, those that are already close to equilibrium), whilst accurately capturing the time-dependent evolution for particles where an implicit solver is required. This approach has also been tested in previous applications with \chimes{} (e.g. \citealt{Rich22raddust}).

\subsection{Multi frequency Radiation Hydrodynamics}
\label{sec:SPHM1RT}

We briefly review the \swiftrt\ \footnote{\swiftrt\ coupled to a hydrogen and helium thermo-chemistry network \citep{Chan21SPHM1RT} is available in the public version of \swift{} (see data availability statement).} radiation hydrodynamics method introduced by \cite{Chan21SPHM1RT}, and discuss its extension to account for
multiple frequencies. \swiftrt\  is a two-moment Lagrangian radiation hydrodynamics method implemented in Smoothed Particle Hydrodynamics (SPH; \citealt{Lucy77SPH,Ging77SPH}). \swiftrt\ therefore benefits from the spatial and temporal adaptabilities of SPH. It evolves two moments of the radiation density and flux densities, together with a closure relation. The computational cost is approximately proportional to the number of gas particles that the radiation propagates through, for a given number of frequency bins and chemical elements; the compute time is nearly independent of the number of sources of radiation. Its adaptivity and the scaling with the number of sources make \swiftrt\ ideal for the typical astrophysical problem, which involves many sources 
and has a large dynamic range in length and time scales. For more details on \swiftrt\, see \cite{Chan21SPHM1RT}.

We extend the two-moment method to multiple frequencies as follows.
Consider the moments of the radiative transfer equation within a frequency bin $i$ (i.e. $[\nu_i,\nu_{i+1}]$). For instance, the radiation energy density in bin $i$ , $E_i$, is given by:
\begin{align}
E_i&=\frac{1}{\tilde{c}}\int_{\nu_i}^{\nu_{i+1}} 4\pi J_\nu{\rm d}\nu\equiv \frac{1}{\tilde{c}}\int_i 4\pi J_\nu{\rm d}\nu,
\label{eq:intIE}
\end{align} 
where $J_\nu$ is the angular-averaged specific intensity, $\nu$ is the frequency, and $\tilde{c}$ is the reduced speed of light (see \S\ref{sec:rsl} below). Similarly, the flux and Eddington tensor within this bin $i$ , ${\bf F}^j_i$ and ${\mathbb P}^{jk}_i$, are given by
\begin{align}
{\bf F}^j_i&=\int_i \hat{\bf n}^j\,4\pi J_\nu{\rm d}\nu,\nonumber\\
{\mathbb P}^{jk}_i&=\frac{1}{\tilde{c}}\int_i \hat{\bf n}^j\,\hat{\bf n}^k\,4\pi J_\nu\,{\rm d}\nu, 
\label{eq:intFP}
\end{align} 
where $\hat{\bf n}$ is the unit vector in the direction in which the radiation travels. We further define the radiation energy per unit gas mass, $\xi_i$, and the radiative flux per unit gas mass, ${\bf f}_i$, both within the bin $i$, as:
\begin{align}
\xi_i\equiv E_i/\rho ;\;\; {\bf f}_i\equiv{\bf F}_i/\rho.
\end{align}

\subsubsection{Moment equations of the radiative transfer equations}
The coupled equations of hydrodynamics and radiative transfer, which
govern the propagation of radiation and its interactions with gas, are \cite[e.g][]{Miha84rhd,Buch83rtff},
\begin{align}
\frac{D\rho}{Dt}+\rho{\bf \nabla\cdot} {\bf v}=0,
\label{eq:masscon}
\end{align}
\begin{align}
\frac{D{\bf v}}{Dt}=-\frac{\nabla p}{\rho}-\nabla\phi+\sum_i\frac{\kappa_i\rho}{\tilde{c}}{\bf f} + {\bf S}_{\bf v},
\label{eq:momcon}
\end{align}
\begin{align}
\frac{Du}{Dt}=-\frac{p}{\rho}{\bf \nabla\cdot\bf v}+\Lambda_{u}+S_{u},
\label{eq:intcon}
\end{align}
\begin{align}
\frac{D\tilde{\xi}_i}{Dt}=-\frac{1}{\rho}\nabla\cdot(\rho{\bf f}_i) -\frac{{\bf\nabla v:} \mathbb{P}}{\rho}+\Lambda_{\xi,i}+S_{\xi,i},
\label{eq:durad}
\end{align}
\begin{align}
\frac{1}{\tilde{c}^2}\frac{D}{Dt}{\bf f}_i=-\frac{{\bf \nabla\cdot \mathbb{P}_i}}{\rho}-\frac{\kappa_i\rho}{\tilde{c}}{\bf f}_i+{\bf S}_{{\bf f},i},
\label{eq:dfrad}
\end{align}
\begin{align}
\mathbb{P}_i=\mathbb{F}_i\tilde{E}_i=\mathbb{F}_i\rho\tilde{\xi}_i\,.
\label{eq:stressrad}
\end{align}
Here, $D/D t$ is the Lagrangian derivative, $\rho$ the gas density, ${\bf v}$ its velocity, $p$ its pressure, and $u$ its thermal energy per unit mass. Furthermore, $\phi$ is the gravitational potential and ${\bf\nabla v:} \mathbb{P}$ is short hand for the contraction $\mathbb{P}^{ij}v_{i,j}$, where $,j$ represents the derivative along direction $j$.

In these equations, $\tilde{c}$ is the \lq reduced\rq\ speed of light, which is a constant fraction of the true speed of light, $c$ (see \S\ref{sec:rsl} for the reduced speed of light approximation). We highlight the quantities affected when using the reduced speed of light approximation with a tilde (e.g., $\tilde{\xi}$). In the original derivation of these equations, $\tilde c=c$.

The right-hand sides represent source and sink terms. The combined heating and cooling rates for gas and radiation are $\Lambda_u$ and $\Lambda_\xi$. $S_i$ is the corresponding source term in direction $\hat{\bf n}^i$. $\kappa$ is the opacity.

The moment method for radiative transfer is very similar to using moments to derive the equations of hydrodynamics from the underlying Boltzmann equation. The rate of change of the zeroth moment of the Boltzmann equation (ie, the density) depends on the first moment (the velocity). The rate of change of velocity depends on the second moment (stress tensor). The hierarchy is closed by introducing an equation of state, which relates density, thermal energy, and pressure. 

For radiation, we see that Eqs.~\ref{eq:durad} and \ref{eq:dfrad} involve the zeroth ($\xi$) and first moment  (${\bf f}$) of the equation for radiative transfer. The time dependence of ${\bf f}$ involves the second moment, the Eddington tensor $\mathbb{P}$. As in the case of hydrodynamics, the hierarchy of moments is closed by imposing a closure relation that relates
$\mathbb{P}$ to $\xi$ and ${\bf f}$. However, this is much harder than in the case of hydrodynamics. The reason is that radiation can stream in a given direction if the gas is optically thin, whereas it can diffuse approximately isotropically if the gas is optically thick. In the first case, the Eddington tensor is far from isotropic, whereas it is isotropic in the second case. The closure relation needs to be able to capture these extreme cases, and interpolate in a meaningful way between them when the gas is neither optically thin nor optically thick. \swiftrt\ uses a modified
\M{} closure relation, which can stabilize radiation fronts in the optically thin limit, see \citealt{Chan21SPHM1RT} for details.

The equations for radiation (Eqs.~\ref{eq:durad} and \ref{eq:dfrad}) are solved with the ``difference'' SPH form \citep{Tric12divBclean}.
As is the case for hydrodynamics, these differential equations can not capture discontinuities. In SPH simulations, this requires implementing artificial \lq viscosity\rq\ terms that allow the code to capture discontinuities such as shocks and contact discontinuities. \swiftrt{} uses novel artificial diffusion and anisotropic flux dissipation terms to capture discontinuities in the radiation field. More details of the implementation can be found in \S 2.5 and 2.6 in \cite{Chan21SPHM1RT}. 

The equations of hydrodynamics (Eqs.~\ref{eq:masscon}, \ref{eq:momcon}, and \ref{eq:intcon}) are solved with the {\sc Sphenix} SPH hydrodynamics method \citep{Borr20SPHENIX}, which is implemented in {\sc swift}. {\sc Sphenix} solves the density-energy equation of motion and is equipped with variable artificial viscosity and conduction. Thus, it can handle fluid mixing and conserve vorticity. Our radiative transfer scheme also works with other SPH methods (e.g., the \lq Gadget-2\rq\ SPH scheme of \citealt{Spri05Gadget2}).

\subsection{The Reduced Speed of Light (RSL) Approximation}
\label{sec:rsl}

Radiation propagates at the speed of light, which is typically {\em much} faster than the speed of an ionization front or any other physical processes (e.g., gas velocities). For example, the physical speed of light ($c\sim 3\times 10^5 ~{\rm km/s}$) is over four orders of magnitude higher than the typical sound speed in the warm interstellar medium ($v_s\sim 10~{\rm km/s}$ for a temperature $\sim 10^4\;{\rm K}$). Hence, the radiative transfer time-step ($\Delta t_c\sim h/c$, with $h$ the smoothing length of a particle) is much shorter than the typical hydrodynamical time-step ($\Delta t_s\sim h/v_s$). This stringent time-step requirement will significantly slow down calculations that include radiation. However, intuitively, it would seem that the speed of the ionization front should be the one limiting the time step, rather than $c$. This is the idea behind the reduced speed of light (RSL) approximation.

\cite{Gned01OTVET} introduced the RSL approximation: replace $c$ by a reduced speed of light, $\tilde c$, in the radiative transfer equation,
\begin{align}
c\rightarrow \tilde{c}; 
\label{eq:rsl}
\end{align}
\begin{align}
n_i\rightarrow \tilde{n}_i,
\label{eq:rsln}
\end{align}
in all equations; here $n_i$ is the photon number density. Under RSL, the radiative transfer time-step can be increased from
$\Delta t_c\sim h/c$ to $\Delta t_c\sim h/\tilde{c}$. As long as $\tilde{c}$ is larger than any characteristic speed in simulations, \cite{Rosd13ramsert} found that RSL correctly captures the progression of an ionization front. \cite{Chan21SPHM1RT} showed that the solutions of photo-ionization equilibrium in a \lq Str\"{o}mgren sphere\rq\ type set-up are also correctly reproduced in the RSL approximation.

We use the RSL approximation (Eq.~\ref{eq:rsl}) in \swifter{} in both the thermo-chemistry  (\S\ref{sec:chemistryeq}) and propagation (\S\ref{sec:SPHM1RT}) step\footnote{It is not correct to 
only use $\tilde c$ in the propagation step but not the thermo-chemistry step, since this will result in unphysical over-ionization, as shown by \cite{Ocvi19DSL}.}. We will examine the robustness of this approximation in the following sections. 

Unfortunately, RSL has several limitations. Firstly, RSL is not suitable when the ionization front moves very fast, which may happen when the optical depth is small. Such a scenario is encountered in cosmological reionization simulations. When using $\tilde c\ll c$, 
RSL overestimates the volume-weighted neutral hydrogen fraction after reionization \citep{Ocvi19DSL}. Second, RSL cannot be applied to systems with high optical depth, when the radiation-diffusion time scale is short \citep{Skin13M1}. Finally, RSL fails if the injection rate of the radiation source decreases fast (for example, when a source of photons is switched off). In this case, RSL underestimates the rate at which the photon density drops, so the gas takes much longer to recombine. Examples of this scenario include the rapid variability (flickering)
of an active galactic nucleus \citep{Oppe18AGN}, or an \ion{H}{II} region where the central star's ionising luminosity has dropped rapidly as the star reached its end of life. We illustrate the latter case in Appendix \ref{sec:issueRSL}.

\subsection{Coupling the radiation equations with thermo-chemistry}
\label{sec:chemistryeq}

The \chimes{} implementation discussed by \cite{Rich14CHIMESI,Rich14CHIMESII} assumed a time-independent background radiation field in the \lq grey\rq\ approximation (i.e. one spectral bin for Extreme-UltraViolet (EUV) photons with energies 13.6 eV - $\infty$ and one for Far-UltraViolet (FUV) photons with energies 6-13.6 eV, with a time-independent amplitude in both bins).

In this study, we extend \chimes{} in two ways. Firstly, we include multi-frequency spectral bins, where the number and energies of the bins can be chosen by the user of the code. Second, we couple the radiation field with the rate equation to account for the loss or gain of photons in each spectral bin, as gas is ionised or recombines, or as photons are absorbed by dust. This enables us to couple \chimes{} to the radiative transfer module\footnote{The additional functionality that we have developed in this paper has been implemented in the public \chimes{} repository.}. In \S\ref{sec:modChimes}, we describe
how we account for multiple spectral bins, the evolution of the radiation field, and diffuse radiation. In S\ref{sec:radfieldeq}, we introduce the evolution equations for radiation energy and flux.

\subsubsection{Thermo-chemistry equations in the presence of multiple spectral bins}
\label{sec:modChimes}

Firstly, we extend \chimes{} to handle multiple spectral bins (with the number of spectral bins $N_{\rm bin}$), given an angular-averaged specific intensity $J_\nu$ (in fact, only the spectrum shape of $J_\nu$ is required). The main difference from \cite{Rich14CHIMESI,Rich14CHIMESII} is that we integrate over discrete energy bins rather than over all energies of an entire spectrum. Our implementation also incorporates the reduced speed of light approximation in \chimes.

In the following, we use $i$ to index a particular spectral bin (e.g. ${\rm bin}_i=[\nu_{i}-\nu_{i+1}]$) and $\texttt{j}$ to represent species (e.g.\ion{H}{I},\ion{He}{I},\ion{He}{II},\ion{O}{II}). $\texttt{j}+{\rm I}$ indicates an extra ionization state, e.g. \ion{He}{I}+{\sc i} = \ion{He}{II}. This term vanishes if the extra ionization state and associated cross section do not exist.

Let $\tilde n_i$ denote the photon number density in bin $i$, 
\begin{align}
\tilde{n}_{{i}}=\frac{1}{\tilde{c}}\int_{\nu_i}^{\nu_{{i}+1}}{\rm d}\nu\,\frac{4\pi J_\nu}{h\nu}\equiv\frac{1}{\tilde{c}}\int_{i}{\rm d}\nu\,\frac{4\pi J_\nu}{h\nu}.
\end{align}
As before, the tilde indicates that the photon density is modified from the physical number density in the RSL approximation. The average energy per photon, $\bar{e}_{i}$, is the radiation energy density divided by the photon number density $n_{{i}}$: 
\begin{align}
\bar{e}_{i} = \frac{\tilde{E}_{i}}{\tilde{n}_{{i}}}=\frac{\int_{i}{\rm d}\nu\,4\pi J_\nu}{\int_{i}{\rm d}\nu\,4\pi J_\nu/h\nu}.
\label{eq:meanphotonenergy}
\end{align}

We consider the photo-ionization of species due to both FUV and EUV radiation (FUV: 6-13.6 eV; EUV: >13.6 eV).
Let $\sigma_\texttt{j}(\nu)$ be the frequency-dependent ionisation cross section of species $\texttt{j}$, with threshold frequency $\nu_{\texttt{j}}$. We use the fits by \citealt{Vern96sigmacross} to compute $\sigma_\texttt{j}(\nu)$. Integrating $\sigma_\texttt{j}(\nu)$ over bin $i$ yields
\begin{align}
\sigma_{{i}\texttt{j}} = \frac{1}{ \tilde{n}_{{i}}\tilde{c}}\int_{i}{\rm d}\nu\,\frac{4\pi J_\nu}{h\nu}\sigma_{\texttt{j}}(\nu).
\label{eq:sigma}
\end{align}
The energy injected into the gas per photo-ionization, averaged over bin $i$, is
\begin{align}
\epsilon_{i\texttt{j}} =&\frac{1}{\sigma_{i\texttt{j}} \tilde{n}_{i}\tilde{c}}\int_i{\rm d}\nu\, \frac{4\pi J_\nu}{h\nu}\sigma_{\texttt{j}}(h\nu-h\nu_{\texttt{j}}).
\label{eq:epsilon}
\end{align}
We note that $\bar{e}_i$, $\sigma_{i\texttt{j}}$ and $\epsilon_{i\texttt{j}}$ depend only on
the {\em shape} of the assumed spectrum within the given bin, not its amplitude.
We also note that $\tilde n\,\tilde c=n\,c$, therefore $\bar{e}_i=\tilde{\bar{e}}_i$, $\sigma_{i\texttt{j}}=\tilde{\sigma}_{i\texttt{j}}$ and $\epsilon_{i\texttt{j}}=\tilde{\epsilon}_{i\texttt{j}}$.

To simplify the calculation, we keep $\bar{e}_i$, $\sigma_{i\texttt{j}}$ and $\epsilon_{i\texttt{j}}$ {\it constant} during a simulation\footnote{This approximation is also adopted in many other astrophysical radiative transfer code, e.g. \citealt{Rosd13ramsert,Pawl11multifRT,Kann19AREPORT}.}. This can be done by fixing the spectrum shape within the spectral bin, although the normalization within a spectral bin is allowed to vary. Hence, we can pre-compute $\bar{e}_i$, $\sigma_{i\texttt{j}}$ and $\epsilon_{i\texttt{j}}$ for all ions
in \chimes{} 

\footnote{We also include the Auger process using the data from \cite{Vern95photocrosssection} and \cite{Kaas93Auger}.}. We provide a {\sc Python} script in the \chimes{} package\footnote{See Data Availability statement.} to compute $e_i$, $\sigma_{i\texttt{j}}$ and $\epsilon_{i\texttt{j}}$ 
given the user-defined spectral bins and the assumed spectral shape $J_\nu$ of sources.

The rate equations for non-molecular species are ordinary differential equations (ODEs) of the form
\begin{align}
\frac{\partial n_\texttt{j}}{\partial t} =&-\sum_i\sigma_{i\texttt{j}}\tilde{c}\tilde{n}_{i}n_\texttt{j}+\sum_i\sigma_{i\texttt{j}-{\rm I}}\tilde{c}\tilde{n}_{i}n_{\texttt{j}-{\rm I}}\nonumber\\
&-\beta_{\texttt{j}}n_en_{\texttt{j}}+\beta_{\texttt{j}-{\rm I}}n_en_{\texttt{j}-{\rm I}}\nonumber\\
&+\alpha_{{\rm A/B},\texttt{j}+I}n_e n_{\texttt{j}+{\rm I}}-\alpha_{{\rm A/B},\texttt{j}}n_e n_{\texttt{j}}\nonumber\\
&+{\rm additional\; processes}.
\label{eq:chem}
\end{align}
The first line describes photo-ionization, the second line collisional ionisation (temperature-dependent collisional ionisation coefficient $\beta$), and the third line recombination (temperature-dependent recombination coefficient $\alpha$); the final line includes any other processes. Recombination can be described either in the case-A (optically thin recombination lines) or case-B (on-the-spot recombinations) approximation.

Our implementation can account
for diffuse recombination radiation, i.e., for the photons emitted by recombining gas in case-A: such photons are added to the 
photon density in the appropriate spectral energy bin.
Alternatively, the user can select to apply the {\it On-The-Spot approximation} (OTS) (e.g. \citealt{Oste06nebulae}), in which case Eq. \ref{eq:chem} uses the case-B recombination coefficient ($\alpha_{\rm B}$). Under OTS, the diffuse radiation is assumed to be immediately reabsorbed in the vicinity (see also \S\ref{sec:radfieldeq}). OTS is often a good approximation for direct recombinations to the ground state of \ion{H}{I} in \ion{H}{II} regions when the optical depth of such photons is high.

Additional processes include, for example, charge transfer (see \citealt{Rich14CHIMESI} for the implementation in \chimes). 
This process can be important in certain scenarios, for example, \cite{Oppe13noneqA} demonstrated that charge transfers are important for \ion{O}{III} in collisionally excited metal-enriched gas.

Eq.~\ref{eq:chem} are subject to two sets of constraints. Firstly, the number density of an element is equal to the sum of the number densities of its ionization states,
\begin{align}
\sum_{\texttt{j}\in {\rm ele}} n_{\texttt{j}}+\sum_{\rm mol}s_{k,{\rm mol}}n_{\rm mol}=n_{\rm ele, tot},
\end{align}
where the sum is over all species of an element and $n_{\rm ele, tot}$ is the total number density of the element (regardless of its ionization state). $s_{k,{\rm mol}}$ is the number of atoms of the element $k$ in molecules. However, in this work, we do not include molecular networks, so the density of molecular species ($n_{\rm mol}$) is zero.

Secondly, the number of free electrons is equal to the overall charge of the ions (overall charge neutrality):
\begin{align}
\sum_{\texttt{j}} N_{e,\texttt{j}}n_{\texttt{j}}=n_e\,,
\end{align}
where $N_{e,\texttt{j}}$ is the charge of ion $\texttt{j}$ (in units of the electron charge). We use these two constraints as independent checks on the accuracy of the {\sc Cvode} solvers, rather than reducing the number of ODEs integrated. We therefore integrate the rate equations for all species and evaluate these constraint equations. If the results from the {\sc Cvode} solvers and the constraint equations differ by more than 1 per cent, we re-normalise the abundances accordingly to ensure that the constraints are adhered to. \cite{Oppe13noneqA} adopted a similar approach.

The photo-heating rate, which enters Eq.~\ref{eq:intcon}, is given by
\begin{align}
\rho S_u = \sum_{i\texttt{j}} \epsilon_{i\texttt{j}}  \sigma_{i\texttt{j}} \tilde{c}\tilde{n}_{i}n_\texttt{j},    
\end{align}
where $\epsilon_{i\texttt{j}}$ is 
given by Eq.~\ref{eq:epsilon}. Other thermal processes include metal line cooling, photoelectric heating, and molecular cooling (see \citealt{Rich14CHIMESI} for details).

The gas temperature is needed when solving
Eqs.~\ref{eq:chem} because the collisional ionisation and the recombination coefficients are temperature dependent. The gas temperature can be calculated from its internal energy $u$, as
\begin{align}
T=(\gamma-1)\frac{\mu m_{\rm H}}{k_{\rm B}}u
\end{align}
where $\mu$ is the mean molecular mass in units of the proton mass, $m_{\rm H}$.

\subsubsection{Evolution of the radiation field}
\label{sec:radfieldeq}
If the number of spectral bins is
$N_{\rm bin}$, we introduce $2\times N_{\rm bin}$ ODEs, half describing the evolution of the radiation energy density and half for the amplitude of the radiation flux.

The photon number density rate equation is 
\begin{align}
\frac{\partial \tilde{n}_{i}}{\partial t} =-\sum_\texttt{j}\sigma_{i\texttt{j}} \tilde{c} \tilde{n}_{i} n_\texttt{j} + \sum_j n_en_{\texttt{j}} (\alpha_{{\rm A},\texttt{j}}-\alpha_{{\rm B},\texttt{j}})+S_\gamma\,.
\label{eq:ngamma}
\end{align}
From left to right, terms are
photon loss due to ionizations, the impact of recombinations, and photon injection by a source, $S_\gamma$.

The term $(\alpha_{{\rm A},\texttt{j}}-\alpha_{{\rm B},\texttt{j}})$ describes the diffuse radiation from the recombination of hydrogen and helium. \swifter{} is able to follow the emission, absorption, and propagation of diffuse recombination radiation (see also Eq.\ref{eq:chem}; for a demonstration, see Fig. \ref{fig:Stromgren3dmfHHe}). If we take the OTS approximation (the diffuse radiation is reabsorbed near the point of creation), this diffuse radiation term will be set to zero. 

We assume that the change in flux caused by opacity is isotropic, 
so that all three components of the photon flux vector are reduced by the same relative factor. Therefore, we only need one ODE per energy bin to evolve the flux vectors, instead of one ODE for each direction. We implement this by 
computing the flux \lq reduction factor\rq, $g_{\gamma,i}(t)$, which
is the relative decrease in the photon flux vector ${\rm f}_{\gamma,i}$:
\begin{align}
{\rm f}_{\gamma,i}(t) = {\rm f}_{\gamma,i}(t=0)g_{\gamma,i}(t)\,;
\end{align}
by definition, $g_{\gamma,i}(t=0)=1$. 
We update ${\rm f}_{\gamma,i}$ with $g_{\gamma,i}$ after each RT time-step.
The rate of change of the flux reduction factor is,
\begin{align}
\frac{\partial g_{\gamma,i}}{\partial t} = -\sum_j\sigma_{i\texttt{j}} \tilde{c} g_{\gamma,i} n_\texttt{j}.
\label{eq:ggamma}
\end{align}

When \chimes{} is coupled to a radiative transfer module in \swifter{}, Eqs.~\ref{eq:ngamma} and \ref{eq:ggamma} are integrated together with the chemical rate equations and thermal equation, using the integration scheme described in \S\ref{sec:modChimes}.

\subsection{The impact of dust on gas and radiation}
\label{sec:dustmethod}
\swifter{} is currently restricted to a simple dust model with a constant dust-to-gas ratio, and a specified nature of the dust particles. We defer considerations of more complicated dust models, which include dust creation and dust destruction, such as described by e.g. \citealt{Chob22dust,Tray25dustISM}, to future work.

\subsubsection{Dust-ion and dust-electron interactions}
\label{sec:dustchem}
Dust chemical processes
include recombinations of ions on grains and photoelectric heating of gas by grains. Ions can recombine with electrons when they are adsorbed on the surface of a dust grain. The efficiency of this process depends on $G_0 \sqrt{T}/n_e$ \citep{Wein01grainrecom}, where $G_0$ is the strength of the FUV radiation in so-called \lq Habing units\rq\ \citep{Habi68}\footnote{$G_0 = 1$ corresponds to the energy density of the interstellar radiation field in the FUV band, 6-13.6 eV, in the solar neighbourhood, as determined by \cite{Habi68}. This corresponds to
an energy density of $5.29\times10^{-14}\; {\rm erg}/{\rm cm}^3$ in this band, and a FUV photon flux of approximately $10^8~{\rm photon~cm}^{-2}{\rm ~s}^{-1}$. The flux depends on the assumed spectral shape, which determines the average energy per photon in this band.}. Photoelectric heating is heating by electrons ejected
from grains by the photoelectric effect.
The efficiency of this process also depends on $G_0 \sqrt{T}/n_e$ \citep{Bake94photoelectric}. 

\swifter{} follows the propagation of the FUV radiation, enabling the calculation of
the amplitude of the FUV energy density in Habing units on each SPH particle. The resulting value is passed to \chimes{}, which will compute the rate of recombinations on dust, and the rate of photoelectric heating by the dust for that particle.

\subsubsection{Scattering and absorption of radiation on dust}
\label{sec:dustatten}

Dust grains can absorb and scatter photons, resulting in dust scattering and attenuation. We restrict our analysis to isotropic processes, governed by the following radiative transfer equation:
\begin{align}
\frac{\mathrm{d} I_\nu}{\mathrm{d} s} = \alpha_\nu\left[\eta_\nu J_\nu -I_\nu\right ]\,.
\label{eq:rtdust}
\end{align}
Here, $I_\nu$ is the specific radiation intensity, $J_\nu$ the angle-averaged specific intensity. $\alpha_\nu$ is the extinction coefficient, which combines
absorption, $\alpha_\nu^{\rm abs}$, and scattering, $\alpha_\nu^{\rm scatt}$,
\begin{align}
\alpha_\nu = \alpha_\nu^{\rm abs}+\alpha_\nu^{\rm scatt},
\end{align}
and $\eta_\nu$ is the albedo,
\begin{align}
\eta_\nu = \frac{\alpha_\nu^{\rm scatt}}{\alpha_\nu^{\rm abs}+\alpha_\nu^{\rm scatt}}.
\end{align} 
The first term on the right-hand side of Eq.~\ref{eq:rtdust} represents isotropic scattering, whereas the second term represents extinction. 

We consider a dust model with a mixture of carbonaceous grains and amorphous silicate grains. Carbonaceous grains are polycyclic aromatic hydrocarbon (PAH)-like when sufficiently small, and graphite-like when large (see e.g. \citealt{Li01dustIR}). The optical properties of the grains are taken from \cite{Drai03dustscatteringI}. We assume that the size distribution of the grains follows the case \lq A\rq\ model of \cite{Wein01dustdis}, with extinction factor $R_V=4.0$ as renormalized by \cite{Drai03dustreview}\footnote{The dust data from various models are available at \href{https://www.astro.princeton.edu/~draine/dust/dustmix.html}{https://www.astro.princeton.edu/~draine/dust/dustmix.html}}.

The moment equations for dust attenuation follow from Eq.~\ref{eq:rtdust} by multiplying with $\hat{n}$ and integrating over direction and frequency,
\begin{align}
\left.\frac{\partial \tilde{n}_i}{\partial t}\right|_{\rm dust} = -\tilde{c}\tilde{n}_i\left(\kappa^{\rm abs}_if_d\rho \right ),
\label{eq:ndust}
\end{align}
\begin{align}
\left.\frac{\partial {\bf f}_i}{\partial t}\right|_{\rm dust} = -\tilde{c}{\bf f}_i\left(\kappa^{\rm ext}_if_d\rho \right ),
\label{eq:Fdust}
\end{align}
where $\kappa^{\rm abs}_i$ and $\kappa^{\rm ext}_i$ are the mean absorption and extinction opacity (per unit dust mass; in ${\rm cm}^2~{\rm g}^{-1}$) within a spectral bin. Here, $f_d$ is the dust-to-gas ratio, which we assume to scale linearly with metallicity. The normalization is set by the Milky Way value of $f_d\approx 0.006$ at solar metallicity.  

We account for dust by reducing both the radiation density and radiative flux
in an operator split fashion: radiation is attenuated in the Radiative Transfer module, as described by Eqs. \ref{eq:ndust} and \ref{eq:Fdust}. The updated radiation densities and fluxes are passed into the \chimes{} thermo-chemistry solver.

\subsection{Direct radiation pressure on dust and gas}
\label{sec:dirradpre}
We account for direct radiation pressure by tracking the changes of 
the momentum flux of radiation before and after the thermo-chemistry step for each gas particle. The change in momentum changes the velocity of the gas particle at a rate
\begin{align}
\left.\frac{D{\bf v}}{Dt}  \right |_{\rm radpre} &= \left.-\frac{1}{\tilde{c}c}\sum_i\frac{\partial {\bf f}_i}{\partial t}\right|_{\rm thermo-chem}
\label{eq:Fdirpre}
\end{align}
In this way, we account for the direct radiation pressure coming from absorption by {\it both} dust and gas.

We currently do not model multi scattered infrared photons, which can boost the momentum injection rate significantly at high optical depth \citep[see, e.g.][]{Ishi15multiscRadforce,Thom16radpre}.

To ensure that the momentum transferred to the gas is independent of the value of $\tilde c$, we have to use $1/(c\tilde{c})$ in Eq.~ \ref{eq:Fdirpre}. However, doing so means that the loss of momentum by radiation is not equal to the gain in momentum by the gas, when $\tilde c\neq c$, i.e., in the RSL approximation. Nevertheless, we will demonstrate that this approximation can give accurate results in standard tests (\S\ref{sec:radpretest}).

\subsection{The injection of radiation by a source}
\label{sec:inject}
We follow the method described by \cite{Chan21SPHM1RT} to inject radiation from a source into the gas. Radiation is injected
by \lq star\rq\ particle $k$ to the surrounding gas particles within its interaction radius with a photon number injection rate $\dot{N}_{k,{\rm inj}}$ (for each frequency bin, $i$). The injection rate to a particular gas particle $l$ is weighted by $1/r_{kl}^2$, where $r_{kl}$ is the distance between gas particle $l$ and star particle $k$ as well as the volume of the gas-particle, $m_l/\rho_l$:
\begin{align}
\Delta N_{kl} = \frac{m_l}{\rho_l r_{kl}^2}\frac{\dot{N}_{k,{\rm inj}}\Delta t_k}{N_{\rm nor}}\,,
\end{align}
where $\Delta t_k$ is the time step of the star, and $N_{\rm nor}$ is a normalization term, such that the total radiation injected by star $k$ in all gas particles combined is $\dot{N}_{k,{\rm inj}}$. We inject the radial radiation flux assuming that the surrounding medium is optically thin.

We convert the photon injection rate $\dot{N}_{\rm inj}$ in [photons/s] to the energy injection rate, $\dot{E}_{\rm inj}$. It is straightforward in the case of a single spectral bin:
\begin{align}
\dot{E}_{\rm inj} =  \dot{N}_{\rm inj}\bar{e}.
\end{align}
If there are multiple spectral bins, the energy injection rate to a spectral bin $i$ is
\begin{align}
\dot{E}_{{\rm inj},i} =\frac{\dot{N}_{\rm inj}\bar{e}_i \tilde{n}_i}{\sum_k \tilde{n}_k}= \frac{\dot{N}_{\rm inj}\bar{e}_i\int_i{\rm d}\nu\,4\pi J_\nu/h\nu}{\sum_n\int_n{\rm d}\nu\,4\pi J_\nu/h\nu},
\end{align}
where $\bar{e}_i$ is calculated from Eq.~\ref{eq:meanphotonenergy}.

\subsection{Implementation in the {\sc Swift } hydrodynamics code}
\label{sec:swiftimplement}
\swift\ is an open-source\footnote{See Data Availability statement.} cosmological simulation code based on SPH hydrodynamics.
Calculations are performed using task-based parallelism to exploit the multiple levels of parallelism when many-core nodes interact in an {\sc mpi} environment \citep{Scha18SWIFTascl,Scha23SWIFT}. 
Excellent strong and weak scaling is achieved using a task-based parallelism, graph-based domain decomposition, cell-based decomposition of the computational volume, and dynamic asynchronous {\sc mpi} communication \citep{Scha16SWIFT,Borr18SWIFT}. Self-gravity is implemented with the fast multipole method, coupled with a particle-mesh algorithm. Recent applications of \swift\ include performing a very large cosmological galaxy formation simulation with cooling, reaching $z=0$ \citep{Scha23FLAMINGO}. \cite{Scha23SWIFT} provides full details
of {\sc swift}.

Task-based parallelism is an efficient parallelization strategy for many-core architectures. The algorithm breaks down the computation into many small tasks, with task dependencies and task conflicts specified at the implementation stage. During a run, independent cores execute individual tasks asynchronously, with a core picking another task from the stack as soon as it completes the current task. Dependencies and conflicts are handled by the locking and unlocking of tasks. The parallelisation strategy aims to minimize load imbalance.  The calculations associated with \swifter{} - radiative transfer and related thermo-chemistry - are incorporated into this framework \citep{Ivko23GEARRT}.

In some non-trivial scenarios, for example, a highly inhomogeneous particle distribution, some of the tasks may have significantly more particles than on average, which may make them computationally expensive. In this case, the task-based parallelism may suffer from load imbalance. This is a significant issue in \swifter{}, because the radiative-transfer thermo-chemistry tasks typically dominate the compute time (see \S\ref{sec:timing}). 
In this study, we alleviate this problem by splitting the radiative-transfer thermo-chemistry tasks into smaller units if the number of particles per task is larger than some number, e.g., 50. These smaller units can be executed independently by different cores: this significantly reduces load imbalance as we demonstrate in \S\ref{sec:timing}.

The radiation time step, $\Delta t_c\approx h/c$, is usually much shorter than other time steps, e.g. the hydrodynamic time step $\Delta t_s\approx h/v_s$ (where $v_s$ is the local sound speed). Rather than performing the whole calculation with the smallest time step, we exploit the fact that when $\Delta t_c\ll \Delta t_s$, we can perform many radiation steps before the density structure of the gas can change. We do so by sub-cycling: performing many steps in which we update the radiation, but not the hydrodynamics.
This radiation sub-cycling concept has successfully been applied in many radiative transfer codes, e.g. \cite{Rosd13ramsert,Kann19AREPORT}. The implementation of the sub-cycling scheme in \swift\ is described by \cite{Ivko23GEARRT}.

\section{Tests of \swifter{}}
\label{sec:verification}
We perform three types of tests to verify the implementation of 
radiative transfer and photo-chemistry in \swifter{}: (1) the photo-ionization of a single gas parcel, which tests the interaction of radiation with chemistry in \chimes{} (\S\ref{sec:iliev0}); (2) direct radiation pressure (\S\ref{sec:radpretest}); (3) the evolution of the abundances in a static \ion{H}{II} region set-up, which tests the combination of radiative transfer and thermo-chemistry  (\S\ref{sec:Stromgren}). In each case, we compare the results of \swifter{} to analytic solutions, if they exist, or against other codes.

The radiative transfer implementation in \swifter{} has been tested
extensively in \cite{Chan21SPHM1RT}. Tests include radiation propagation in the optically thin limit, the trapping of an ionization front by a dense clump, and the evolution of an \ion{H}{II} region with radiation hydrodynamics.

\subsection{Test 1: thermo-chemistry in a single gas particle}
\label{sec:iliev0}

\begin{figure}
\includegraphics[width=0.48\textwidth]{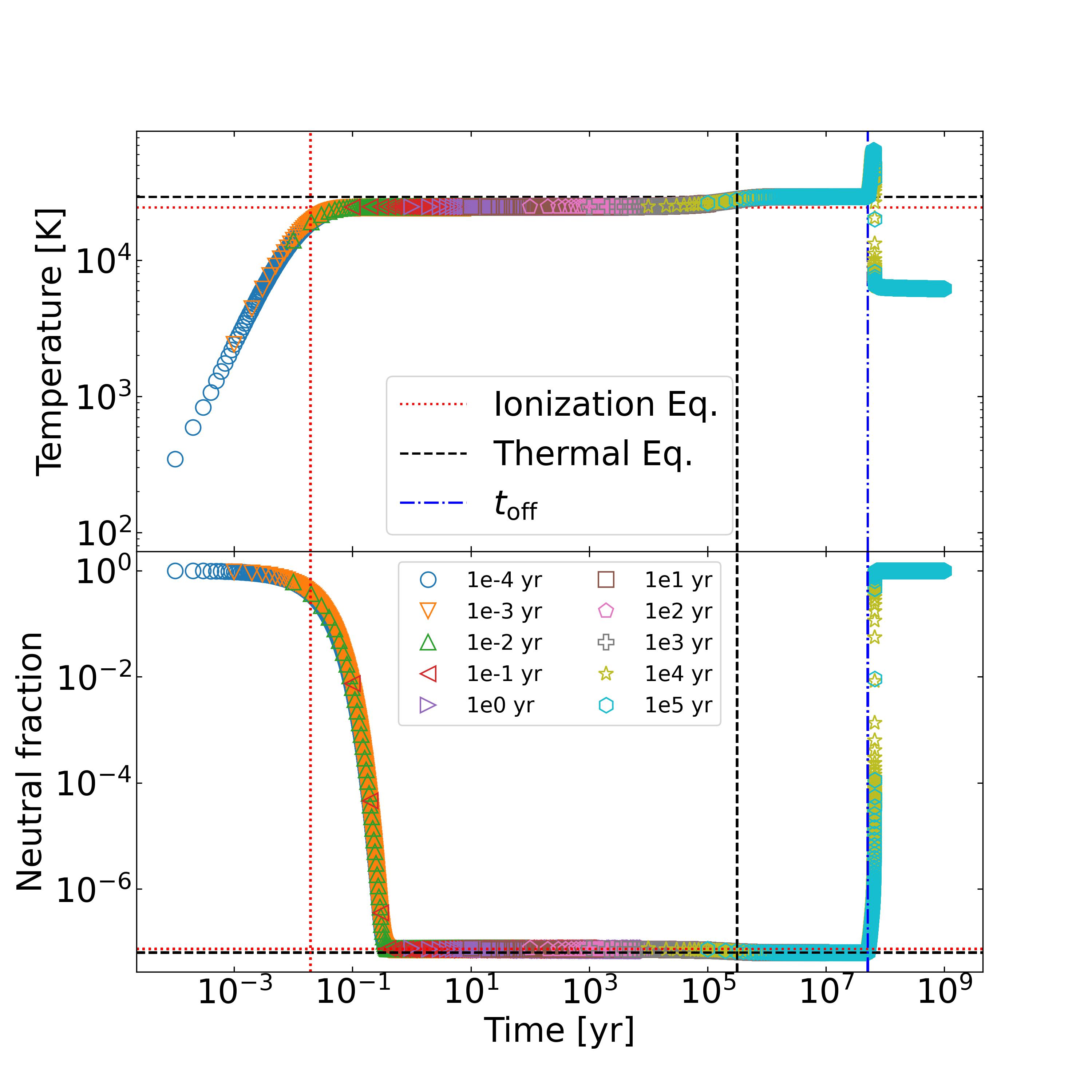}
\vspace{-7mm}
\caption{Test 1: Photo-heating of a single gas parcel with multi-frequency radiation (\S\ref{sec:iliev0}). Hydrogen gas is ionized by radiation with a $10^5\;{\rm K}$ blackbody spectrum with constant photon flux in the optically thin limit, the source is switched off at time $t=t_{\rm off}$ ({\em vertical dot-dashed blue lines}). {\em Symbols}
correspond to results obtained with the \chimes{} implementation in \swifter{}, where {\em different colours} corresponding to selecting different output time intervals in \chimes{}. The different symbols fall on top of each other, demonstrating that the 
\chimes{} results are independent of the chosen output time intervals. The {\em vertical dotted red lines} indicate the ionisation time scale, after which the gas is in ionization equilibrium ({\em horizontal dotted red lines}). After a recombination time ({\em vertical black dashed line}), the gas reaches thermal equilibrium ({\em horizontal dashed black lines}). The \swifter{} results match well with the analytical expectations (see text).}
\label{fig:HIcooling3}
\end{figure}

\begin{figure}
\includegraphics[width=0.48\textwidth]{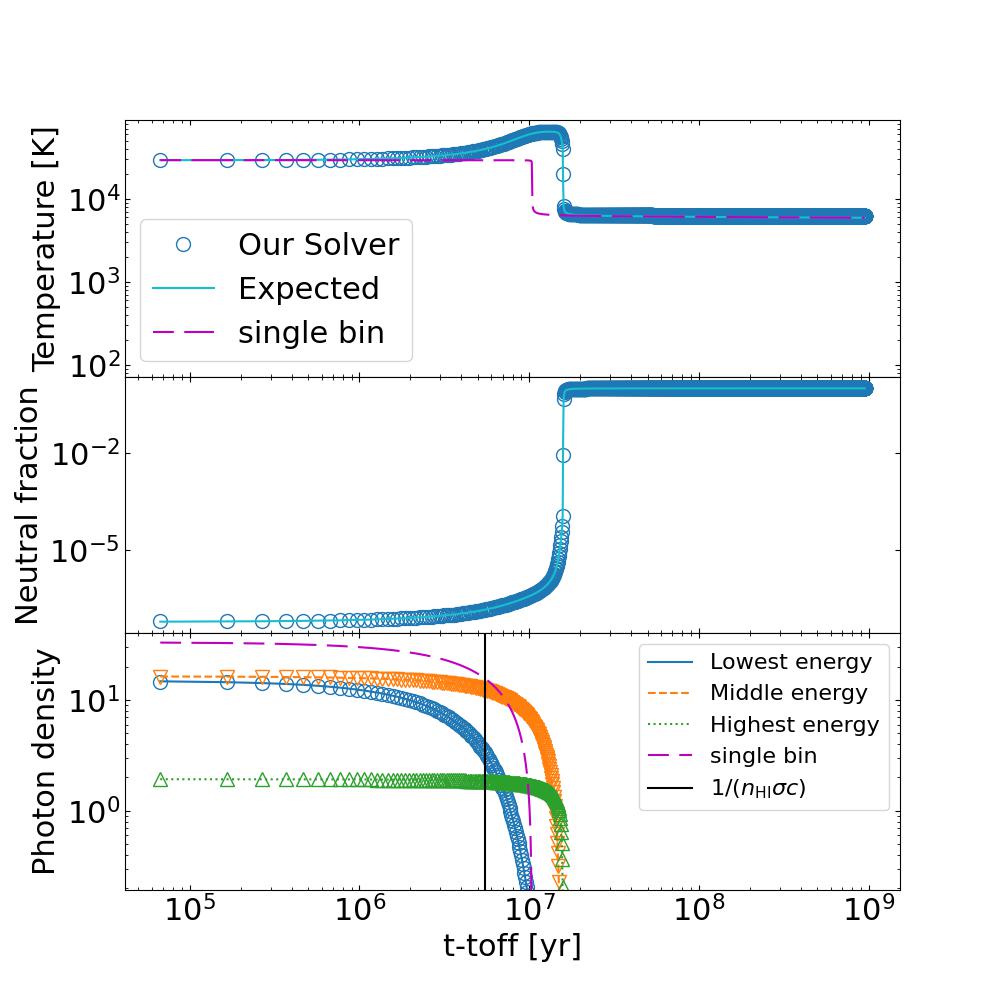}
\vspace{-7mm}
\caption{Test 1: Gas and photon evolution after the radiation source is turned off (\S\ref{sec:iliev0}). The horizontal axis is the time measured from the source switched off, $t-t_{\rm off}$. {\em Blue symbols} are the result from \swifter{} using three spectral energy bins, the {\em cyan solid line} is obtained from integrating the rate equations using the {\sc odeint} routine in {\sc scipy}. There is excellent agreement between these different codes. The {\em dashed magenta line} in the top panel is the \swifter{} result when only one, rather than three, spectral bins are used in the calculation. 
When using a single bin, the temperature does not increase briefly after source turn-off.
The photon density plot in the lower panel illuminates the physics of spectral hardening, which causes the temperature increase in the first case. As the gas starts to recombine, the photon density decreases. However, the energy density in the lowest energy bin ({\em blue symbols}) drops faster than in the highest energy bin ({\em green symbols}). This spectral hardening causes an increase in temperature. The {\em dashed magenta line} shows the photon density in the case of only one spectral bin for completeness.}
\label{fig:HIcooling3_toff}
\end{figure}

In this test, an optically thin gas particle is photo-heated by multi-frequency UV radiation, and we examine the evolution of the ionic abundances and temperature of the particle. This test is similar to test 0 (Part 3) performed in the cosmological radiative transfer comparison project \citep{Ilie06RTcom}, as well as the thermo-chemistry test I in \cite{Chan21SPHM1RT}. However, \cite{Chan21SPHM1RT} used a single frequency bin, whereas here, we consider three frequency bins: [13.6-24.6, 24.6-54.4, 54.4 $\rightarrow\infty$] eV (corresponding to the ionisation energies of \ion{H}{I}, \ion{He}{I}, \ion{He}{II}, respectively). This test assesses the accuracy of the thermo-chemistry solver. The radiation field is held constant, so the result is independent
of any radiative transfer effects.

The details of the test are as follows. A parcel of gas is composed of hydrogen
with density $n_{\rm H}=1~{\rm cm}^{-3}$, initial neutral fraction $x_0=1$, and initial temperature $T_0=100~{\rm K}$. At time $t=0\;{\rm yr}$, the parcel is being irradiated with a black body spectrum of temperature $T=10^5\;{\rm K}$ and a fixed photon flux of $F_{\rm photon} = 10^{12} \;{\rm photons}\,{\rm s}^{-1}{\rm cm}^{-2}$. The source is switched off after a time $t_{\rm off}=5\times 10^7 \;{\rm yr}$ and we follow the evolution until time $t_e=10^9 \;{\rm yr}$. We take ${\tilde c}=c$ (i.e., without RSL). We do not account for elements other than hydrogen, dust, or cosmic ray physics. The evolution of the neutral fraction and the gas temperature is plotted in 
Fig.~\ref{fig:HIcooling3}, and can be understood as follows.

\subsubsection{Evolution before source turn-off}
When the source is switched on, the gas is photo-ionized and photo-heated, and reaches  ionization equilibrium after the ionization timescale $\tau_i$, given by
\begin{align}
\tau_i&\equiv \frac{1}{\sum c\sigma_{ i, {\rm HI}} n_i}\approx10^{-2.3}\;{\rm yr}\,,
\label{eq:ionizationtimescale}
\end{align}
and is plotted a the vertical dotted red line in Fig.~\ref{fig:HIcooling3}. 
$\tau_i$ is the timescale to ionize around half of the gas if there is only photo-ionization, which is a good approximation initially since other processes (in particular, recombination) are much slower.

The neutral fraction $1-x_{\rm ion}$ in ionization equilibrium follows from balancing photo-ionisations with recombinations,
\begin{align}
1-x_{\rm ion}=\frac{\tau_i}{\tau_r}\approx\frac{n_{\rm H}\alpha_B}{\sum_i c\sigma_{i, {\rm HI}} n_i}\approx10^{-7.4}\,,
\label{eq:ionizationx}
\end{align}
where $\tau_i/\tau_r$ is the ratio between the ionization and recombination time-scale, $\tau_r$. The temperature at the ionization equilibrium follows from computing the net amount of energy transferred from the radiation field to the gas to ionize it
to the level $1-x_{\rm ion}$,
\begin{align}
T_{\rm ion}\approx \frac{1}{3k_{\rm B}}\frac{\sum_i\epsilon_{i, {\rm HI}}\sigma_{ i, {\rm HI}}n_i}{\sum_i\sigma_{ i, {\rm HI}}n_i}\approx10^{4.39}\;{\rm K}\,.
\label{eq:ionizationT}
\end{align}
More details are provided in Section 2.8.3 of \cite{Chan21SPHM1RT}.
$1-x_{\rm ion}$ and $T_{\rm ion}$ are plotted as dotted red lines in Fig.~\ref{fig:HIcooling3}.  

After reaching ionization equilibrium, the gas temperature continues to rise until heating balances cooling, i.e., until the gas reaches thermal equilibrium. The time scale to reach thermal equilibrium is approximately the recombination time \citep[see, e.g.][]{Chan21SPHM1RT}, and is plotted as a vertical dashed black line in Fig.~\ref{fig:HIcooling3}.  We solve numerically for the temperature and neutral fraction at which the heating rate is equal to the cooling rate, and plot the corresponding neutral fraction and temperature as horizontal dashed black lines in Fig.~\ref{fig:HIcooling3}.

The simulation predictions for these values are independent of the choice of spectral bins since the photo-heating and photo-ionization rates are not affected by which bins are chosen. This can be seen, e.g., by summing the denominator of Eq.~\ref{eq:ionizationx} using Eq.~\ref{eq:sigma}. However, we will see that the choice of bins affects the evolution {\em after} the source is switched off.
More generally, the choice of spectral bins will affect the simulation results when the radiation propagates and when non-equilibrium effects are taken into account. For more examples of this, see below.

The time evolution of temperature and neutral fraction obtained from \swifter{} are shown in Fig. \ref{fig:HIcooling3}. \chimes{} calls {\sc Cvode} to integrate a system of thermo-chemistry ODEs across an output time interval (see \S\ref{sec:CHIMES_method})\footnote{  We note that {\sc Cvode} employs its own sub-cycling time-steps internally according to the stiffness of the problems.}. First, we examine whether our solver obtains the correct solution even with a coarse time interval. It is essential for thermo-chemistry since the time scales of different stages are very disparate (from $10^{-3}-10^{8}\; {\rm yr}$). In Fig. \ref{fig:HIcooling3}, we plot \swifter's solution for different \chimes{} output time intervals using different colours. All of the simulation points lie on top of each other - demonstrating that the result is independent of the choice of output intervals, as should be the case. The plot also shows that the gas in \swifter{} reaches the correct ionization equilibrium, both in terms of neutral fraction and gas temperature (red dotted line). After a recombination time, the gas in the simulation settles into the correct thermal equilibrium state, both in terms of neutral fraction and of gas temperature (black dashed line). The excellent agreement with the analytical results demonstrates that thermo-chemistry in \chimes{} is accurate and insensitive to the output time intervals.

\subsubsection{Evolution after source turn-off}
\label{sec:test0off}

The source is turned off at $t_{\rm off}=5\times 10^7 \;{\rm yr}$, after which the remaining radiation is consumed by neutral hydrogen atoms. Ionized hydrogen continues to recombine until the neutral fraction returns to close to unity, as shown in Fig.~\ref{fig:HIcooling3}. In the absence of photo-heating, the gas temperature drops to $\sim 8000 \;{\rm K}$, a quasi-equilibrium temperature below which the hydrogen cooling is inefficient.

The gas temperature briefly {\em increases} just after the source is turned off at time $t_{\rm off}$ ($t_{\rm off}$ is indicated by a vertical dot-dashed line). We find that this feature disappears when the calculation is repeated with a single frequency bin
(as was the case in the tests of \citealt{Chan21SPHM1RT,Ilie06RTcom}). We recall that we see this feature in our run, which uses three frequency bins.
To investigate this phenomenon in more detail and verify our solver, we plot $t-t_{\rm off}$ against gas/photon properties in Fig.~\ref{fig:HIcooling3_toff}. We also compare the numerical solution against the expected results.

We compute the thermal evolution by numerically integrating the hydrogen ionization equation (Eqs. \ref{eq:chem} and \ref{eq:ngamma}) with {\sc odeint} in the {\sc scipy} {\sc Python} package \citep{Virt20scipy}. In this integration, we take the initial temperature, neutral fraction, and radiation density from the simulation at $t_{\rm off}$. We also take the collisional ionization rates, recombination rates, and cross-sections from \chimes{}, neglecting other processes (since they are not important in the pure hydrogen case). The numerical solution is plotted with solid lines in Fig.~\ref {fig:HIcooling3_toff}, and agrees very well with the simulated solution.
The excellent agreement demonstrates that our solver is accurate and that the temperature increase following source switch-off is physical.

We can qualitatively understand the features in Fig.~\ref{fig:HIcooling3_toff},
including the temperature increase, as follows. When there is no source, photons are consumed by neutral hydrogen at a rate $\sim 1/(x_{\rm HI}\sigma c)$ (Eq.~\ref{eq:ngamma}; and the vertical line in the lowest panel). Since the photo-ionization cross-section $\sigma$ decreases with increasing photon energy, the lower energy photons are consumed faster than the higher energy photons. Therefore, the gas is photo-heated by increasingly more energetic photons as time progresses past source turn-off. As a result, the gas temperature briefly increases to up to a factor of two until all of the photons are used up, after which the gas returns to the quasi-equilibrium temperature of $T\sim 8000\;{\rm K}$. Spectral hardening is clear from the different rates at which photons with different energies are consumed in the lower panel of
Fig.~\ref{fig:HIcooling3_toff}.

We also plot the single-frequency solution obtained from \swifter{} in Fig.~\ref{fig:HIcooling3_toff} as magenta long-dashed lines. The top panel demonstrates
that in this case, the gas temperature does not increase after the source is turned off. Instead, the gas cools down to $\sim 8000\;{\rm K}$ around $t-t_{\rm off}=10^7\;{\rm Myr}$. The spectral shape remains invariant in this single frequency approximation, and hence the heating rate is simply proportional to the photo-ionization rate, which decreases during recombination. Furthermore, the temperature in the single frequency case drops faster, since the resulting photo-ionization cross-section is higher than those of the middle/highest energy bins in the multi-frequency scenario (as illustrated in the bottom panel of Fig. \ref{fig:HIcooling3_toff}). 

The increase in temperature after source turn-off is a result of spectral hardening and cannot be captured in the single-bin tests of \cite{Ilie06RTcom} or \cite{Chan21SPHM1RT}. We examined the case where the gas is not pure hydrogen but rather a mixture of hydrogen, helium, and other elements.
We find that the increase in temperature following source turn-off is much smaller, which we attribute to the higher cooling rate when more elements are present.

In Appendix \ref{sec:multibin_test}, we repeat Test 1 with different numbers of spectral bins. We find that three spectral bins (we used here) are already sufficient to capture these multi-frequency effects.

\subsection{Test 2: Direct radiation pressure}
\label{sec:radpretest}
\begin{figure}
\includegraphics[width=0.48\textwidth]{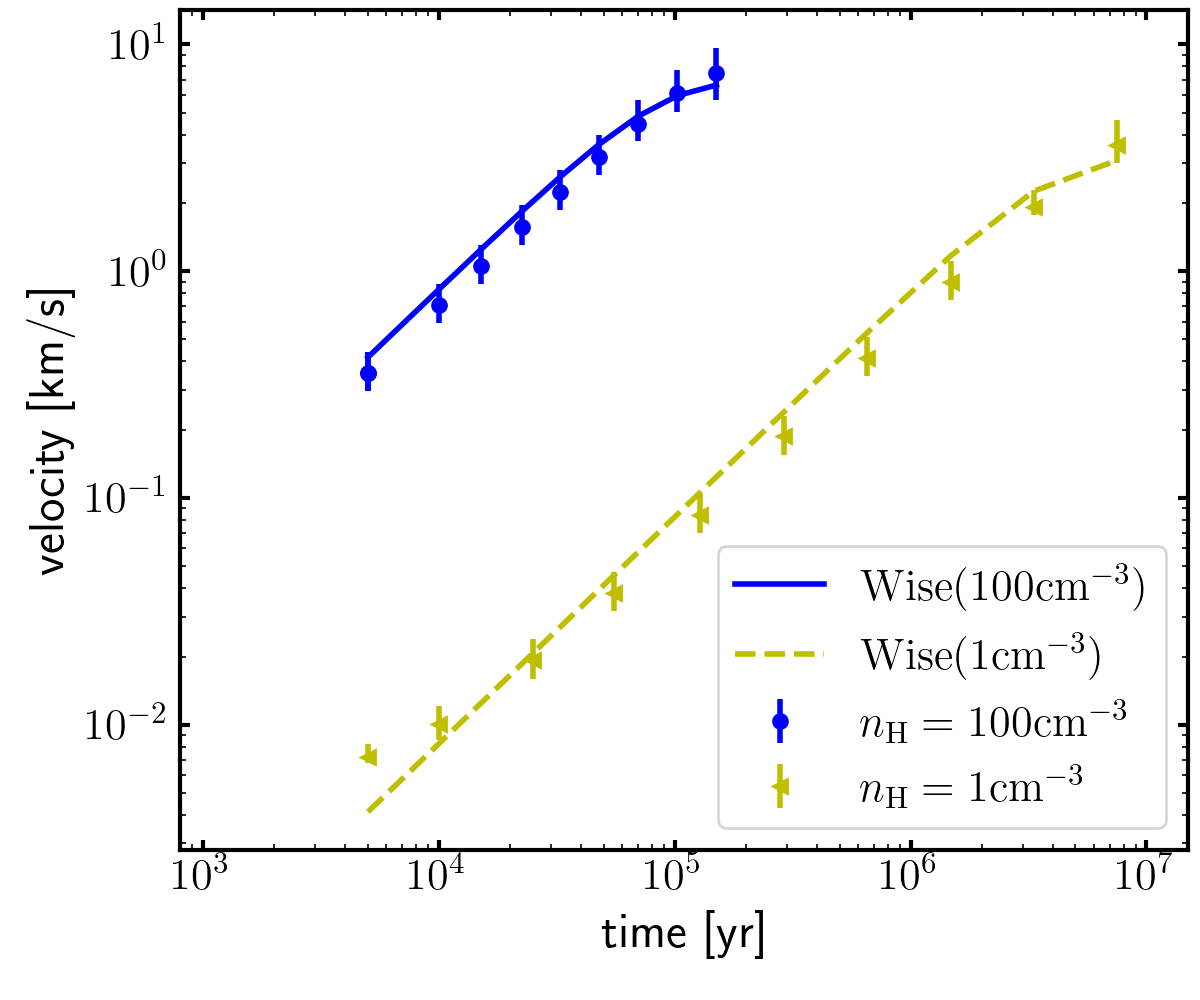}
\vspace{-7mm}
\caption{Test 2: Direct radiation pressure sweeping up gas.
The mean velocity of the ionised gas is shown as a function of time since the radiation source is on. The {\em blue} ({\em yellow}) symbols correspond to an initial density of 100 ${\rm cm}^{-3}$ (1 ${\rm cm}^{-3}$), error bars encompass the 10 and 90$^{\rm th}$ percentiles of radial velocity for particles in the ionised shell. {\sm Solid lines}
show the analytic expectation of Eq.~\ref{eq:wisedirpre}. The simulation agrees very well with the analytic expectation.}
\label{fig:delta}
\end{figure}

We test our implementation of direct radiation pressure described in \S\ref{sec:dirradpre} by performing the test presented by \cite{Sale14dirpre}. Hydrogen gas is initially neutral and uniform in density and temperature. At time $t=0$, 
a source switches on, isotropically emitting photons with energy 13.6~eV at a constant rate. When a hydrogen atom is ionised, it absorbs the momentum of the photon, but is not heated since 13.6~eV is the ionisation energy of hydrogen.
The temperature will nevertheless change due to the change in mean molecular weight per particle. The effect of the radiation pressure is that the gas will start to stream away from the source. The velocity of the gas can be estimated and compared to the result obtained by \swifter{}. We do not account for elements other than hydrogen, dust, or cosmic ray physics.

\begin{figure*}
\includegraphics[width=0.95\textwidth]{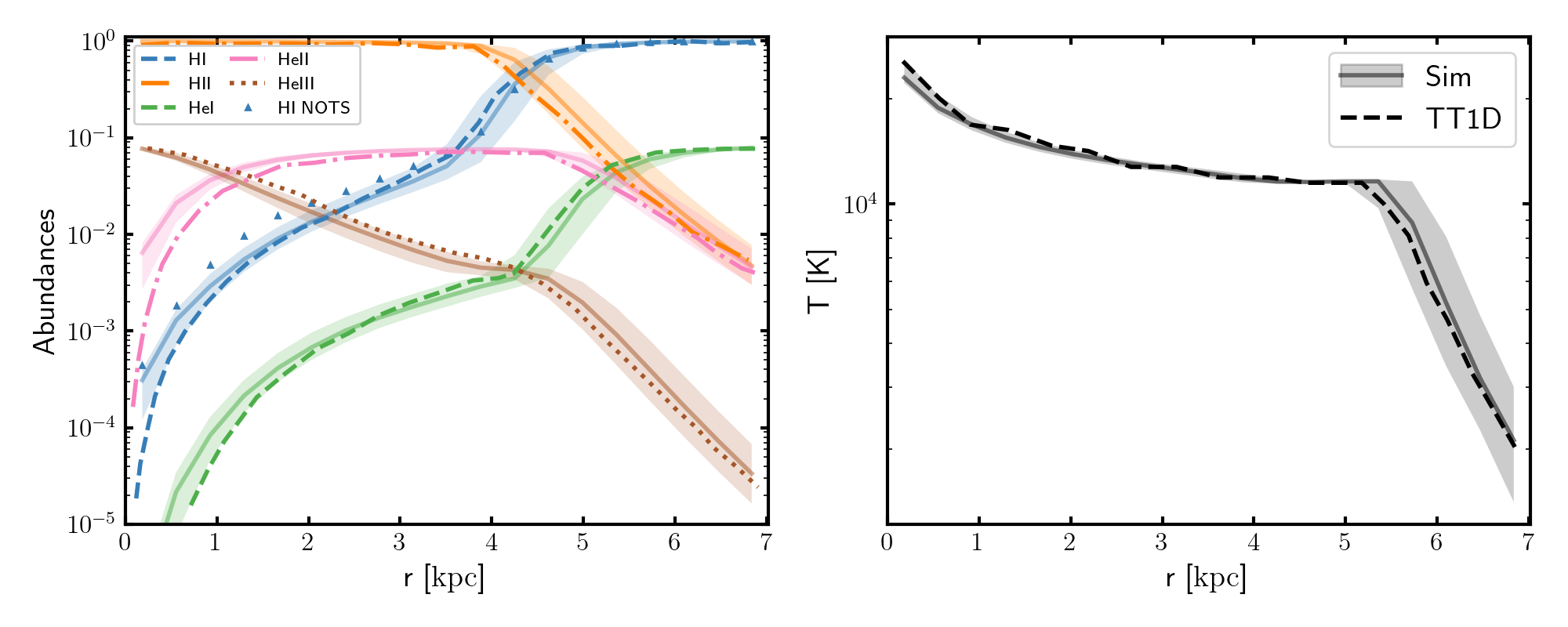}
\vspace{-5mm}
\caption{Test 3: static variable temperature multi-frequency Str\"{o}mgren sphere with hydrogen and helium. The structure of the gas is shown at a time $t=100\;{\rm Myr}$ after the source is switched on. {\it Left panel}: species abundances (relative to hydrogen density $n_{\rm H}$) as a function of radius for different species as per the legend. The median abundances from \swifter{} are shown by {\em solid lines}, the shaded area encompasses the 10-90$^{\rm th}$ percentiles. {\em Dashed or dotted lines} are the results from {\sc tt1d}, taken from \protect\citep{Pawl11multifRT}.{\em Blue triangles} are the abundances for \ion{H}{I} from \swifter{} assuming case-A recombination, which accounts for the emission of ionising photons by recombining gas (the other curves assume case-B, on-the-spot recombinations; see \S\ref{sec:modChimes} and \ref{sec:radfieldeq}). {\it Right panel}: temperature of the gas as a function of radius. The median temperature for \swifter{} is shown as a {\em solid line}, with 10-90$^{\rm th}$ percentiles encompassed by the grey shaded region. Results from {\sc tt1d} are shown as a {\em dashed black line}. The two simulations agree well. Deviations near the source are mostly caused by limitations in the way radiation is injected in \swifter{}.}
\label{fig:Stromgren3dmfHHe}
\end{figure*}

\subsubsection{Analytic solution}
An (approximate) analytic solution to this test case was derived by \cite{Wise12radpre} as follows. Firstly, we assume that all of the ionized gas collects in an optically thick shell with a time-dependent mass $M_{\rm shell}$ and with a momentum $Lt/c$, where $L$ is the source luminosity. We further assume that the optically thick shell initially forms at radius the Str\"{o}mgren radius $r_i$.

The velocity of this shell can now be calculated by expressing momentum conservation,
\begin{align}
v(t) = \frac{Lt}{cM_{\rm shell}} = \frac{Lt}{c}\left [ 4\pi\int_0^rr'^2\rho(r') {\rm d}r'\right ]^{-1}.
\end{align}

If the initial density is constant , i.e. $\rho(r')=\rho_0$, then the solution is
\begin{align}
r(t)&=\left ( r_i^4+2At^2 \right )^{1/4}\nonumber\\
v(t)&=tA\left ( r_i^4+2At^2 \right )^{-3/4},
\label{eq:wisedirpre}
\end{align}
where $A={(3L)}/{(4\pi c\rho_0)}$.

\subsubsection{Numerical simulations}
We perform two simulations, taking the initial hydrogen density to be either $100\;{\rm cm}^{-3}$ or $1\;{\rm cm}^{-3}$. In both cases, the initial temperature is $100\;{\rm K}$. The simulation volume contains $32^3$ gas particles, initially glass-like distributed. The luminosity of the star is $L=10^6L_{\odot }$ (corresponding to an ionisation rate of $\dot{N}_\gamma\sim 1.8\times10^{50}\;{\rm photon~s}^{-1}$). These simulations use the RSL approximation, setting $\tilde{c}=0.01c$.
Fig.~\ref{fig:delta} compares the \swifter{} simulation results to the analytic solution
of Eq.~\ref{eq:wisedirpre}. 

Fig. \ref{fig:delta} shows that the \swifter{} simulation results agree very well 
with the analytic result of Eq.~\ref{eq:wisedirpre}. We note that only ionized gas is accelerated, therefore we only consider the mean radial velocity of ionized gas with neutral fraction $<10^{-4}$.

The slowdown seen most clearly in the blue line at a time $t\sim10^5\;{\rm yr}$, is a result of entrainment, where gas removed from the inner regions starts to accumulate and slows down the expanding shell. The analytic solution does not capture entrainment, as discussed by \citealt{Sale14dirpre}, therefore the simulation and the analytical result do not need to agree at late times, which is why we stop the simulation.

\subsection{Test 3: a multi-frequency source creating an HII region}
\label{sec:Stromgren}

We examine the accuracy of \swifter{} through several multi-frequency \ion{H}{II} region tests. Where possible, we compare our simulation results with previous published work or against {\sc cloudy} calculations.

\subsubsection{Multi-frequency \ion{H}{II} region with hydrogen and helium}
\label{sec:HII}
In this test, we examine the ability of \swifter{} to capture non-equilibrium thermo-chemistry and multi-frequency effects, when a uniform and initially neutral gas is ionised by a source with a black body spectrum. We compare to the results of \cite{Pawl11multifRT}, who
examined the formation of a cosmological \ion{H}{II} region. The density of the gas is therefore low,
hydrogen density of $\sim 10^{-3}\;{\rm cm}^{-3}$. 
This test neglects hydrodynamical and cosmological evolution and focuses on capturing the evolution of the ionised fraction for hydrogen and helium and the temperature of the gas when the source is a black body.

$64^3$ SPH gas particles with a glass-like distribution
fill a cubic computational volume of extent 22-kpc on a side. The composition of the gas is 75 per cent hydrogen and 25 per cent helium by mass, both initially neutral. The density distribution is (and remains) uniform, with a hydrogen number density of $10^{-3}\;{\rm cm}^{-3}$ and initial temperature of 100~K.

At time $t=0$, a central source starts emitting 
photons with a $10^5$~K blackbody spectrum at a constant rate of $5\times10^{48}$ ionizing photons per second.
We capture the shape of the spectrum with five frequency bins, at [13.6-24.6, 24.6-35.5, 35.5-54.4, 54.4-75.0, 75.0-$\infty$] eV\footnote{We adopt the same frequency bins as \cite{Pawl11multifRT} to make a fair comparison. Notice that the ionisation energies of \ion{H}{I}, \ion{He}{I} and \ion{He}{II} are
13.6, 24.6, and 54.4~eV.}. Dust, cosmic rays, and molecular physics are turned off. We also make the RSL approximation, setting $\tilde{c}=0.01c$, and use case-B
recombination.

We compare the simulation result from \swifter{} at 100 Myr against those using the spherically symmetric (1D) multi-frequency radiative transfer code {\sc tt1d}\footnote{The {\sc tt1d} code has been verified against an analytic multi-frequency solution in \cite{Pawl11multifRT}, as well as to results obtained with the {\sc cloudy} 
photo-ionization code (version 08.00, \citealt{Ferl98CLOUDY90}).} as published by \cite{Pawl11multifRT} in Fig.~\ref{fig:Stromgren3dmfHHe}.

The overall agreement is excellent despite the codes using slightly different thermochemistry coefficients. The relative differences in abundances are around 10 per cent except around the centre and close to the ionization front (at $\approx 5$~kpc). The relative temperature difference is within 5 per cent, but is larger near the ionization front. There are small deviations around the centre because we inject radiation in a sphere with radius 1~kpc so that the ionisation fronts are approximately spherical, notwithstanding the relatively small number of SPH particles that receive radiation from the source. The ionization front also propagates slightly faster in \swifter{} because of the extent of the injection region. 

\swifter{} is also able to account for diffuse radiation - ionising radiation emitted by the recombining gas. We can therefore run this problem assuming case-A recombinations without making the on-the-spot approximation. To demonstrate this, we simulate with the identical setup as above but assuming case-A recombinations. The resulting \ion{H}{I} profile is overplotted with triangle markers in Fig.~\ref{fig:Stromgren3dmfHHe}. 

The hydrogen gas in the case-A run is slightly more neutral close to the source, because more ionising photons can cross this region without being lost to recombinations. Nevertheless, the location of the ionisation front is quite similar in both cases.

\subsubsection{Multi-frequency \ion{H}{II} region with hydrogen, helium, and metals}
\label{sec:HIImetal}
\begin{table}
\caption{Abundances of elements by mass for the solar composition from \protect\cite{Wier09abundance}.
These are used in the multi-frequency \ion{H}{II} region test (\S\ref{sec:HIImetal}) and the ionization front propagation simulation (\S\ref{sec:application}).}
\vspace{-3mm}
\begin{center}
    \begin{tabular}{llll}
    \hline
    \hline
H     & 0.706                & Mg    & $5.91\times 10^{-4}$ \\
He    & 0.281                & Si    & $6.83\times 10^{-4}$ \\
C     & $2.06\times 10^{-3}$ &  S & $3.18\times 10^{-4}$\\
N     & $8.36\times 10^{-4}$ &  Ca & $6.59\times 10^{-5}$\\
O     & $5.49\times 10^{-3}$ &  Fe    & $1.10\times 10^{-3}$\\
 Ne    & $1.41\times 10^{-3}$&        &\\
\hline
    \hline
\end{tabular}
\end{center}

\label{table:eleab}
\end{table}

\begin{figure}

\includegraphics[width=0.48\textwidth]{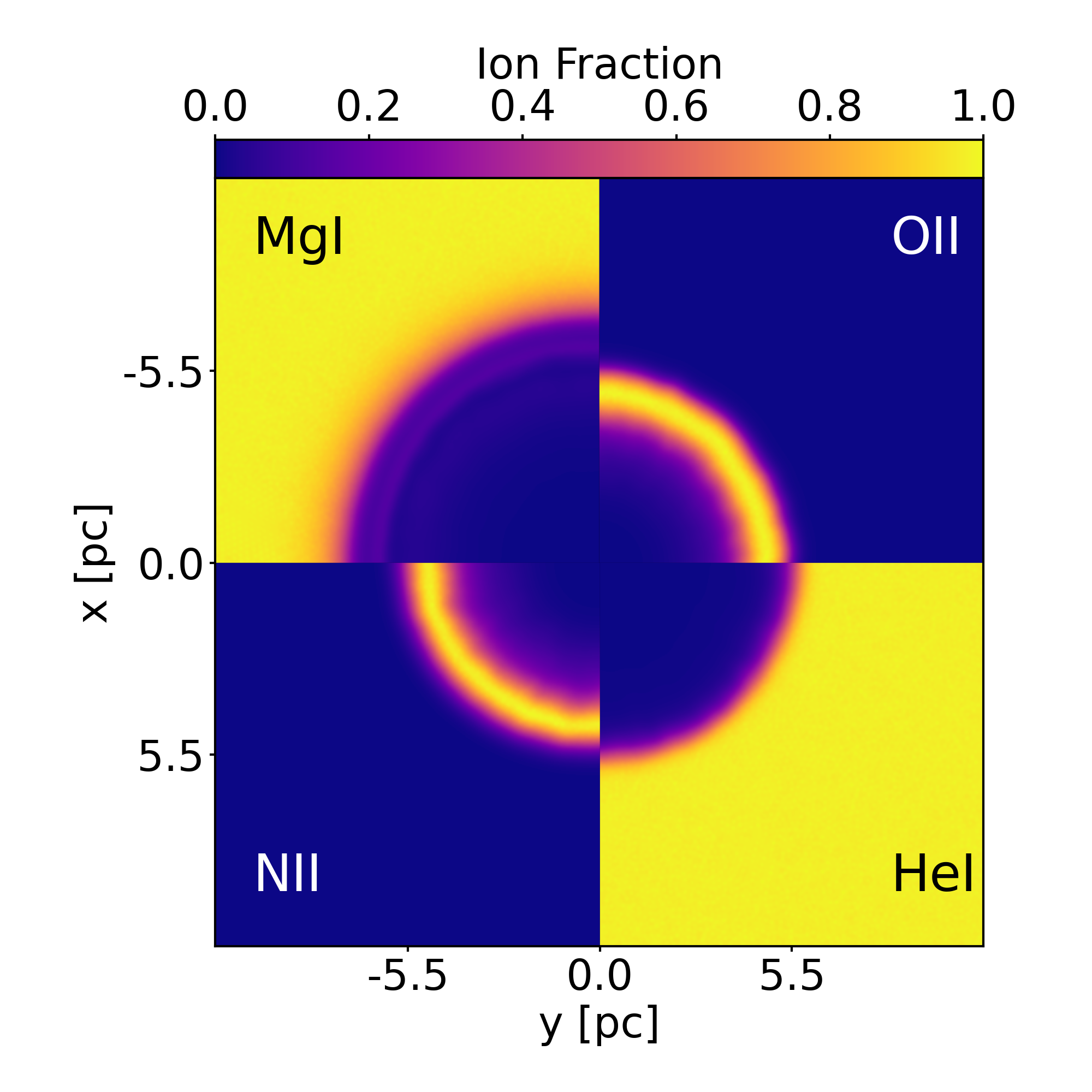}
\vspace{-6mm}
\caption{Test 3: cut through the \ion{H}{II} region
showing ionisation fractions computed by \swifter{} 
at time $t=1.2~{\rm Myr}$ (\S\ref{sec:HIImetal}). Clockwise from top left, the species are
\ion{Mg}{I}, \ion{O}{II}, \ion{He}{I} and \ion{N}{II}.
The slice goes through the centre of the box,
where the source is located.}
\label{fig:Stromgren3dslicelinear}
\end{figure}

\begin{figure*} 
\includegraphics[width=0.98\textwidth]{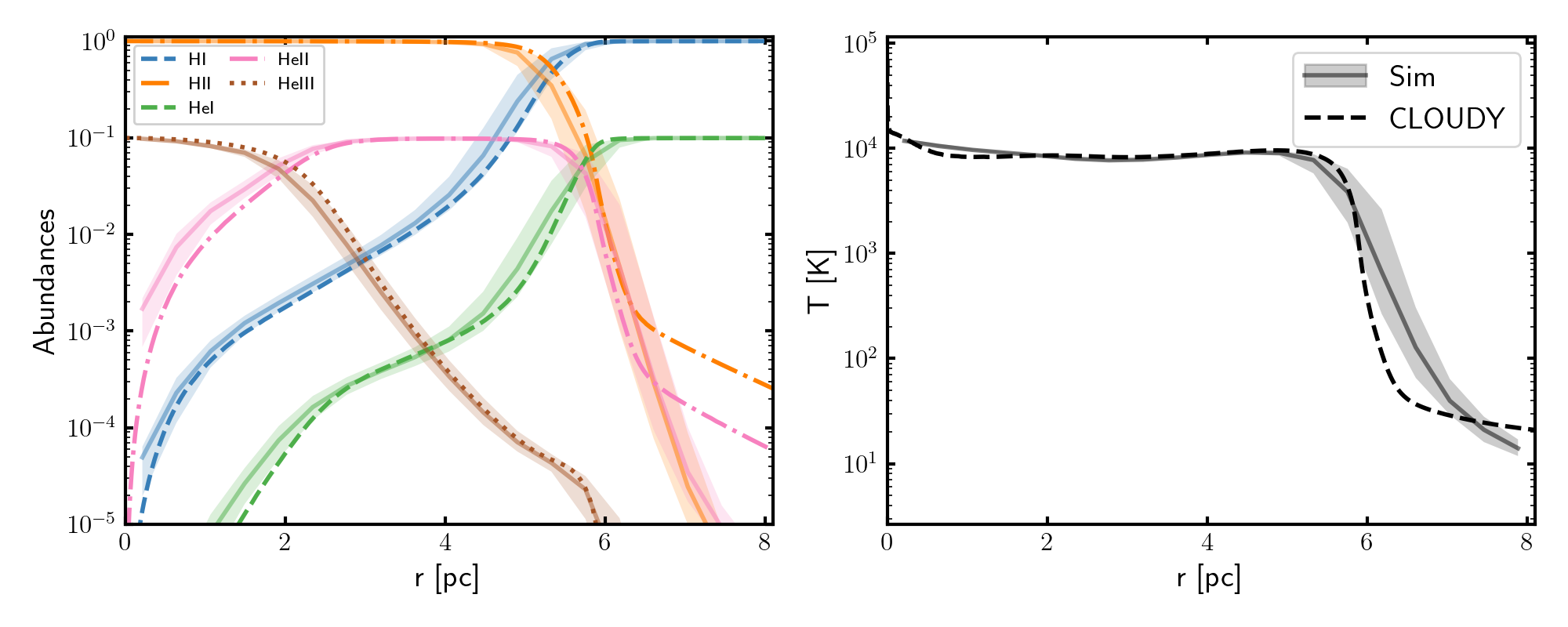}
\vspace{-5.5mm}
\caption{Test 3: as Fig.~\ref{fig:Stromgren3dmfHHe},
but for an \ion{H}{II} region with the solar composition from Table~\ref{table:eleab}. The system is shown at a time $t=1.28\;{\rm Myr}$. {\it Left panel}: species abundance as a function of radius, {\it Right panel}: gas temperature as a function of radius. Results from \swifter{} are shown by a solid line with 10-90$^{\rm th}$ percentiles encompassed by the grey shaded region. Results from {\sc cloudy} are plotted with dashed or dotted lines. 
The results from \swifter{} agree well with those from {\sc cloudy}, except in the very inner region (due to photon injection in the simulation), and near the Str\"omgren radius ($\approx 6$~pc), where the species abundances and the gas temperature are not yet in thermal equilibrium.}

\label{fig:Stromgren3dmfHHeZ}
\end{figure*}

\begin{figure*} 
\includegraphics[width=0.98\textwidth]{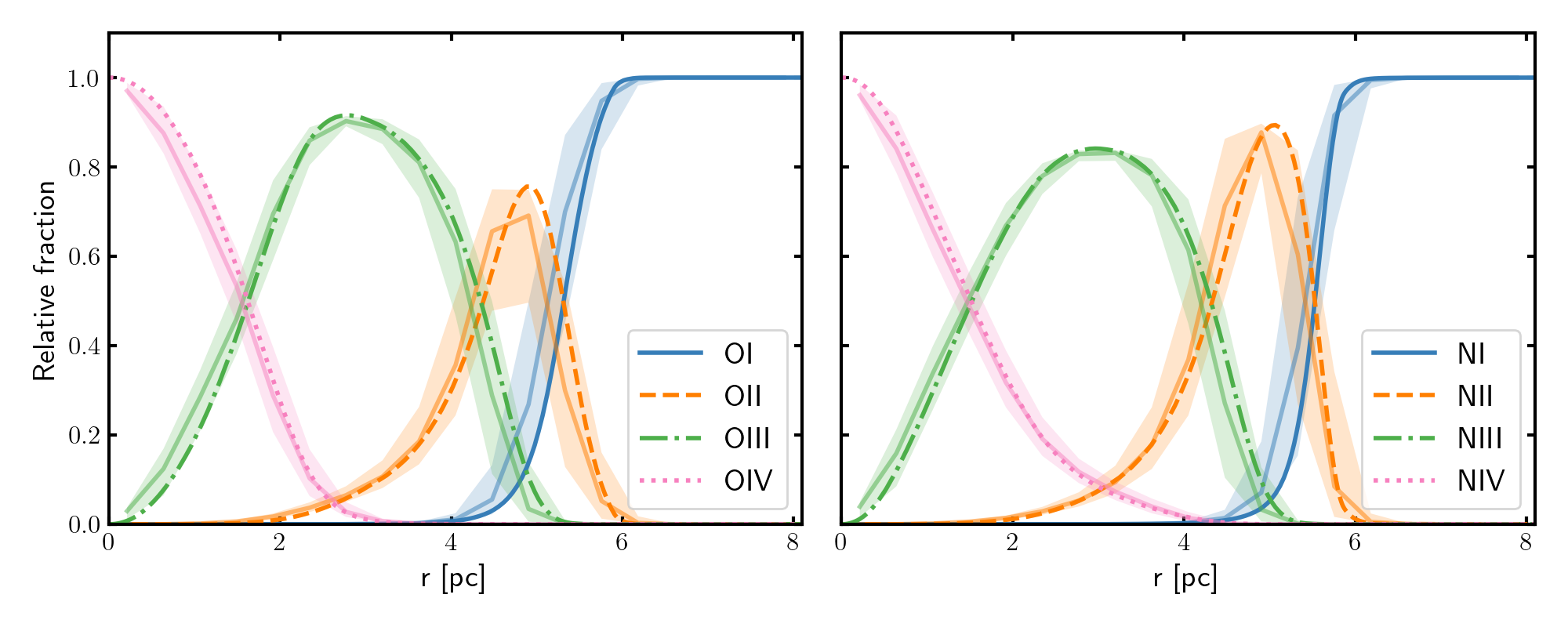}
\vspace{-6mm}
\caption{Test 3: the ionisation fractions of oxygen (left panel) and nitrogen (right panel) (i.e. $n_{\rm OII}/n_{\rm O}$) at a time $t=1.28\;{\rm Myr}$. Results from \swifter{} are shown by a solid line with 10-90$^{\rm th}$ percentiles encompassed by the grey shaded region. Results from {\sc cloudy} are plotted with dashed or dotted lines.  Ionisation fractions from \swifter{} agree well with those of {\sc cloudy}, except in the very inner region (due to photon injection in the simulation), and near the Str\"omgren radius ($\approx 6$~pc), where the ionisation fractions and the gas temperature are not yet in thermal equilibrium.}
\label{fig:Stromgren3dmfHHeZ_metal}
\end{figure*}

In this test, we examine the abilities of \swifter{} to handle metal thermo-chemistry. We place a central source with a blackbody spectrum of temperature $1\times10^5\;{\rm K}$ with an ionising photon rate ($5\times 10^{47} \;{\rm s}^{-1}$). The simulation is run using eight frequency bins with edges [13.6, 18.0, 24.59, 35.5, 54.42, 68.02, 81.62, 95.22, $\infty$] eV. 13.6, 24.6, and 54.4 eV represent the ionisation energies of \ion{H}{I}, \ion{He}{I} and \ion{He}{II}. We include additional bins to sample the energy range between and above these thresholds.


We adopt a uniform gas density of $n_{\rm H} = 10\;{\rm cm}^{-3}$ in a 22 pc box with 128$^3$ gas particles. The elemental composition of the gas is set to the \lq solar abundances\rq\ of Table \ref{table:eleab}, taken from \cite{Wier09abundance}. Notice that \swifter{} follows the ionisation levels of all positive ions of these elements, and in addition tracks ${\rm H}^-$, ${\rm C}^-$, and ${\rm O}^-$.

As before, we turn off dust, cosmic rays, and molecular physics. We also assume case-B recombinations and
make the RSL approximation, with $\tilde{c}=0.01c$. The simulation is run for 1.2~Myr after source turn on, at which point the ionization front has approximately reached the \ion{H}{I} Str\"{o}mgren radius.

We plot the fractions of \ion{Mg}{I}, \ion{He}{I}, \ion{O}{II} and \ion{N}{II} at time $=1.2$~Myr since the source is on, in the form of a slice through the spherical \ion{H}{II} region in Fig.~\ref{fig:Stromgren3dslicelinear}.
The spherical symmetry of the system is approximately preserved, albeit with small deviations, because of the finite spatial resolution and the somewhat irregular distribution of the SPH particles. As expected, low ions, such as \ion{He}{I} and \ion{Mg}{I} are most abundant close to and outside the Str\"omgren radius, whereas intermediate ions, such as, \ion{O}{II} and \ion{N}{II}, are most abundant in a shell, with higher ionisation states dominant inside the shell, and lower ionisation states dominant outside the shell.

More quantitative results are plotted in Figs.~\ref{fig:Stromgren3dmfHHeZ} and \ref{fig:Stromgren3dmfHHeZ_metal}, which have
similar formats to Fig.~\ref{fig:Stromgren3dmfHHe}: the median species abundances from \swifter{} are plotted as solid lines, with the shading encompassing the 10-90$^{\rm th}$ percentiles; different colours correspond to different ions, as per the legend. The dashed and dotted lines in the panels are the equilibrium abundances computed using {\sc cloudy} (version c22.02, last described by \citealt{Ferl17cloudy}). Notice that the \swifter{} results do not assume equilibrium abundances.

Fig.~\ref{fig:Stromgren3dmfHHeZ} compares the hydrogen and helium species abundances and the temperature profile at time $t=1.2$~Myr  against {\sc cloudy}. 

The overall agreement between the simulations is very good within $r<6~{\rm pc}$. The relative differences in abundance and temperature are around 10~per cent, except in the very inner region and close to the Str\"omgren radius ($r\approx 6~$pc). The deviations near the centre
are caused by the extent of the injection radius (we inject over four gas smoothing lengths), as was also the case in the previous test. Deviations near the 
Str\"omgren radius are due to the non-equilibrium effects: species abundances have not yet reached equilibrium at $t=1.2$~Myr in the outer region ($r>6~{\rm pc}$). Using the ionization rates from \chimes{}, we estimate that the time-scale to reach equilibrium at 8~pc is more than 1~Gyr - much longer than the simulation time.

In Fig. \ref{fig:Stromgren3dmfHHeZ_metal}, we compare the oxygen and nitrogen ionisation fraction profiles from \swifter{} to {\sc cloudy}. Similar to Fig.~\ref {fig:Stromgren3dmfHHeZ}, the agreement is very good apart from small deviations at the outer radius caused by deviations from equilibrium.

\section{EXAMPLE APPLICATION: PROPAGATION OF AN IONIZATION FRONT INTO THE INTERSTELLAR MEDIUM}
\label{sec:application}
As a realistic example to illustrate the possibilities of \swifter{}, we investigate the propagation of an ionization front in a small patch ($1\;{\rm pc}^3$) of the interstellar medium. In particular, we compare cases where the gas is uniform versus non-uniform and turbulent, and contrast the ionisation fraction computed in equilibrium versus the full non-equilibrium case. The examples allow us to illustrate the impact of density inhomogeneities, FUV/EUV radiation, radiation hydrodynamics, radiation pressure, non-equilibrium evolution of all metal ions available, self-gravity, as well as dust physics. We also include low-energy cosmic ray heating and ionization, as implemented in \chimes{} \citep{Rich14CHIMESI}. In the second part of this section, we process the simulation output and create mock images.

The simulation volume is cubic, with side length $ 1~{\rm pc}$, and it contains $128^3$ gas particles with average hydrogen number density $\langle n_{\rm H}\rangle= 50~{\rm cm}^{-3}$ - therefore the particle mass is $\approx 7\times 10^{-7}{\rm M}_\odot$. The composition
uses the solar abundances from Table \ref{table:eleab}. The gas is initially neutral with a temperature of 15~ {\rm K}. We include dust with a constant dust-to-gas mass ratio of 0.0062565, which is the Milky Way value from \cite{Math77sizegrain, Ferl13}.

At times $t>0$, the cube is irradiated from one side (the positive $x$-direction) with plane-parallel radiation. The constant \ion{H}{I} ionization rate is $1.8\times 10^{-16} {\rm s}^{-1}$, consistent with the mean value in the Milky Way as determined by several surveys \citep[see, e.g.][]{Pado20lecr}. The rates for the other species are then scaled from this value according to the ratios in the UMIST database \citep{LeTe99UMIST}. We impose constant radiation energy and flux injection for all gas particles with $0~{\rm pc}<\,x\,<\;0.2~{\rm pc}$. To avoid reflection, we continuously reduce the radiation energy and flux in the gas particles with $0.8~{\rm pc}<\,x\,<\;1~{\rm pc}$. We impose periodic boundary conditions in the $y$ and $z$ directions.

The injected photon flux for ionising radiation (energy > 13.6 eV) is $\sim 2\times 10^{10}~{\rm cm}^{-2}{\rm s}^{-1}$. Its spectrum is that of a black body with a temperature of $3\times10^4~{\rm K}$. This value is approximately the flux $\sim 1\; {\rm pc}$ away from a massive O/B star \citep{McLe20NGC300} or a few ${\rm pc}$ away from a star cluster. We consider four frequency bins with edges [6.0, 13.6, 24.6, 54.4, $\infty$] eV. The 6-13.6 eV bin represents the FUV radiation, whereas 13.6, 24.6, and 54.4 eV represent the \ion{H}{I}, \ion{He}{I}, and \ion{He}{II} ionization energies, respectively. We 
perform the calculation assuming case-B recombination 
and turn off molecular physics.

\begin{figure*}
\includegraphics[width=0.47\textwidth]{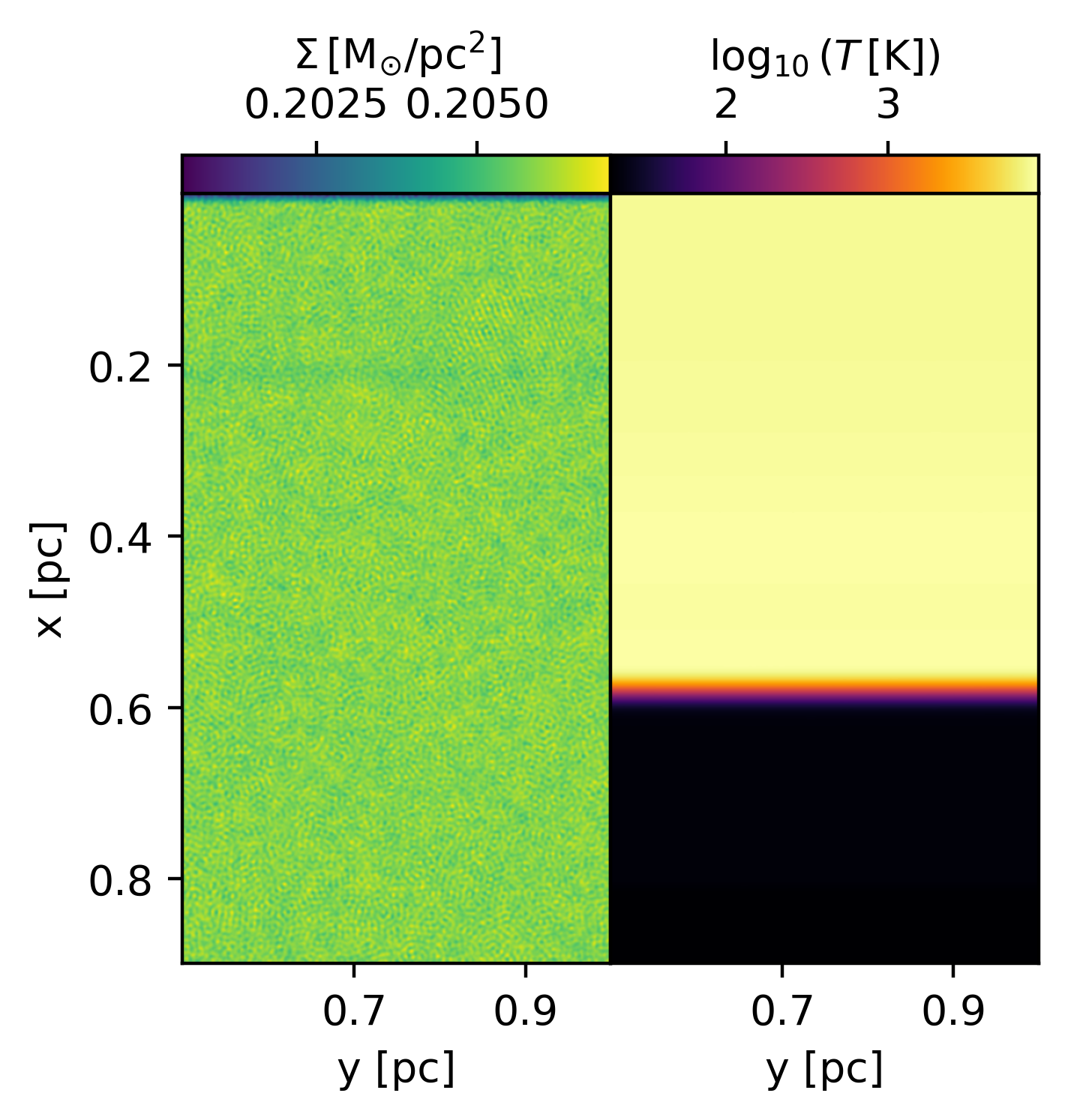}
\includegraphics[width=0.47\textwidth]{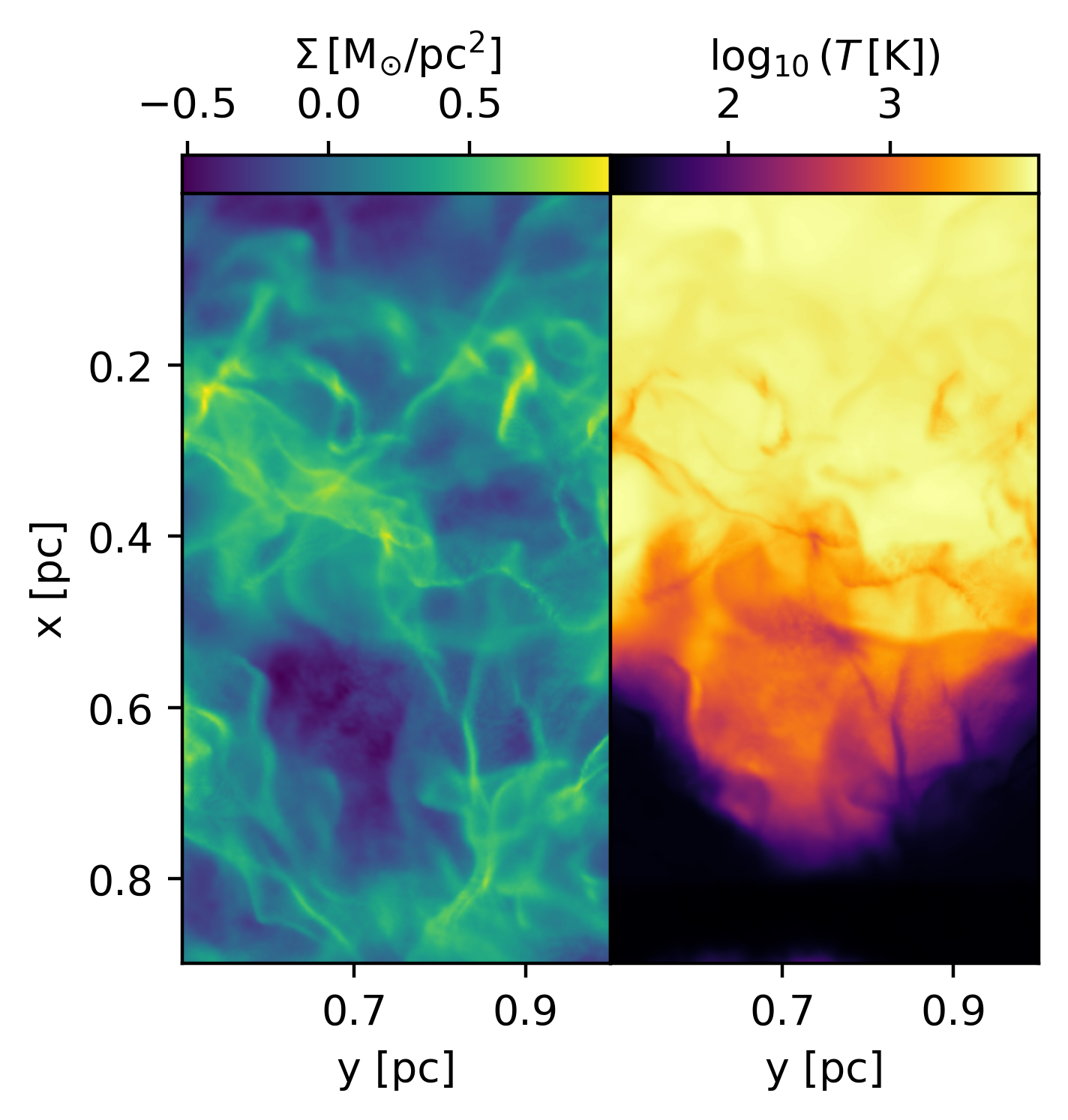}
\vspace{-3.5mm}
\caption{Projected density $\Sigma$ ({\it left subplots}) and mass-weighted temperature $T$ ({\it right subplots}) for the case of initially uniform ({\it left panel}) and initially inhomogeneous ({\it right panel}) density fields. Plane-parallel radiation enters from the top and streams down; the simulation is shown at time $t = 200$~yr from when the radiation is switched on. We project along the $z$-axis and we cut off the areas with $x > 0.9$~pc, where we remove radiation. The colour scales for $\Sigma$ are different in the left and right panels in the surface density plot, to better bring out the contrast in both cases.}
\label{fig:RadBound_rhoT}
\end{figure*}

We perform two simulations: one with a uniform density distribution and another with an inhomogeneous density distribution. In the uniform case, gas particles have a glass-like distribution initially. In the inhomogeneous case, we generate the particle distribution from the isothermal solenoidal turbulence with a Mach number of five, using the {\sc phantom} SPH code described by \cite{Pric18phantom}. {\sc phantom} drives turbulence via an Ornstein-Uhlenbeck process (first suggested by \citealt{Eswa88OUturb}), which was implemented in SPH by \cite{Pric10}. We drove turbulence over five crossing times. The resulting particle distribution is taken as our initial condition for the inhomogeneous simulation. We restrict ourselves to using the inhomogeneous density distribution resulting from turbulence, but we do not drive turbulence during the \swifter{} run. We also set the initial gas velocities to zero.

\begin{figure*}
\includegraphics[width=0.49\textwidth]{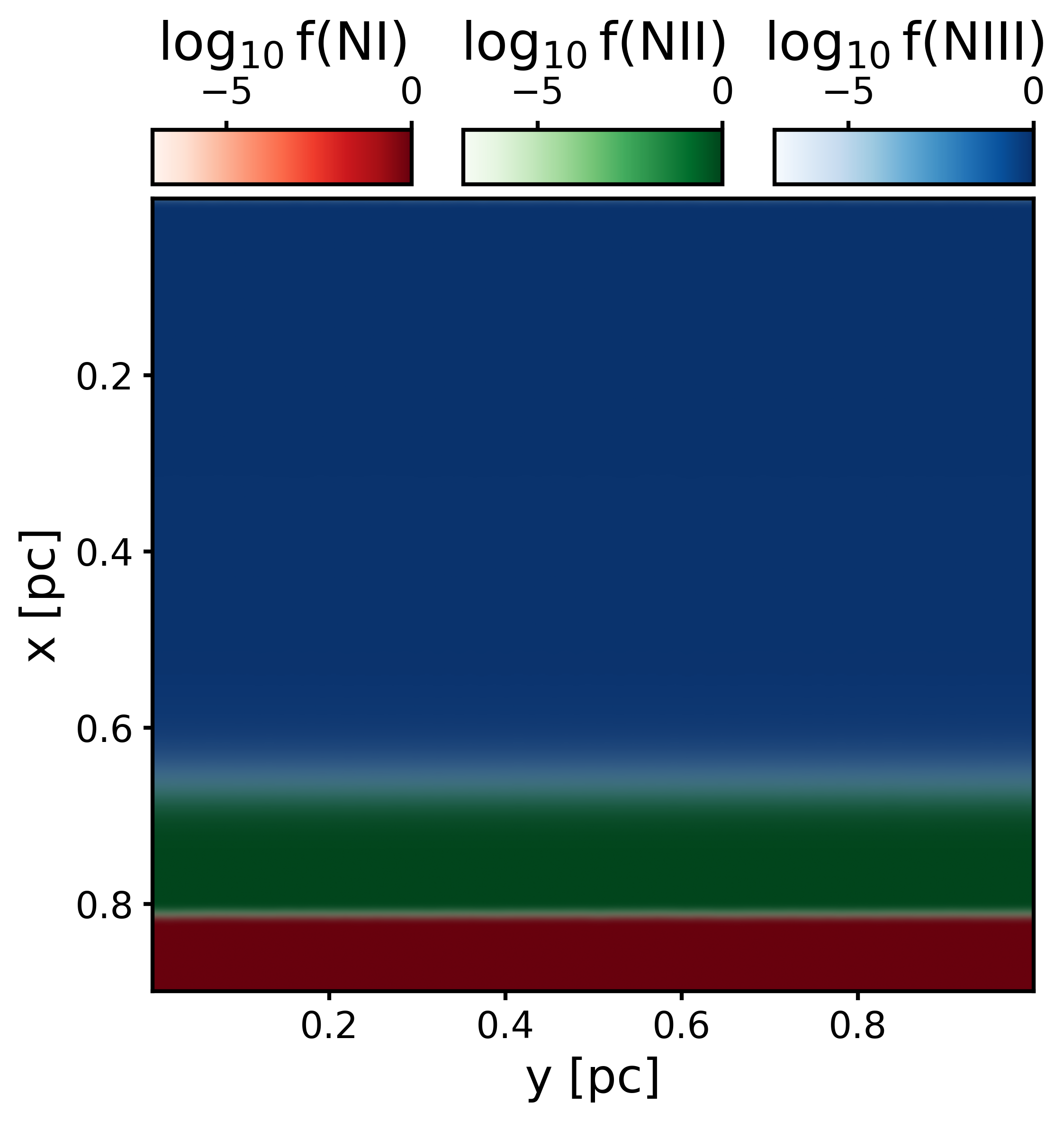}
\includegraphics[width=0.49\textwidth]{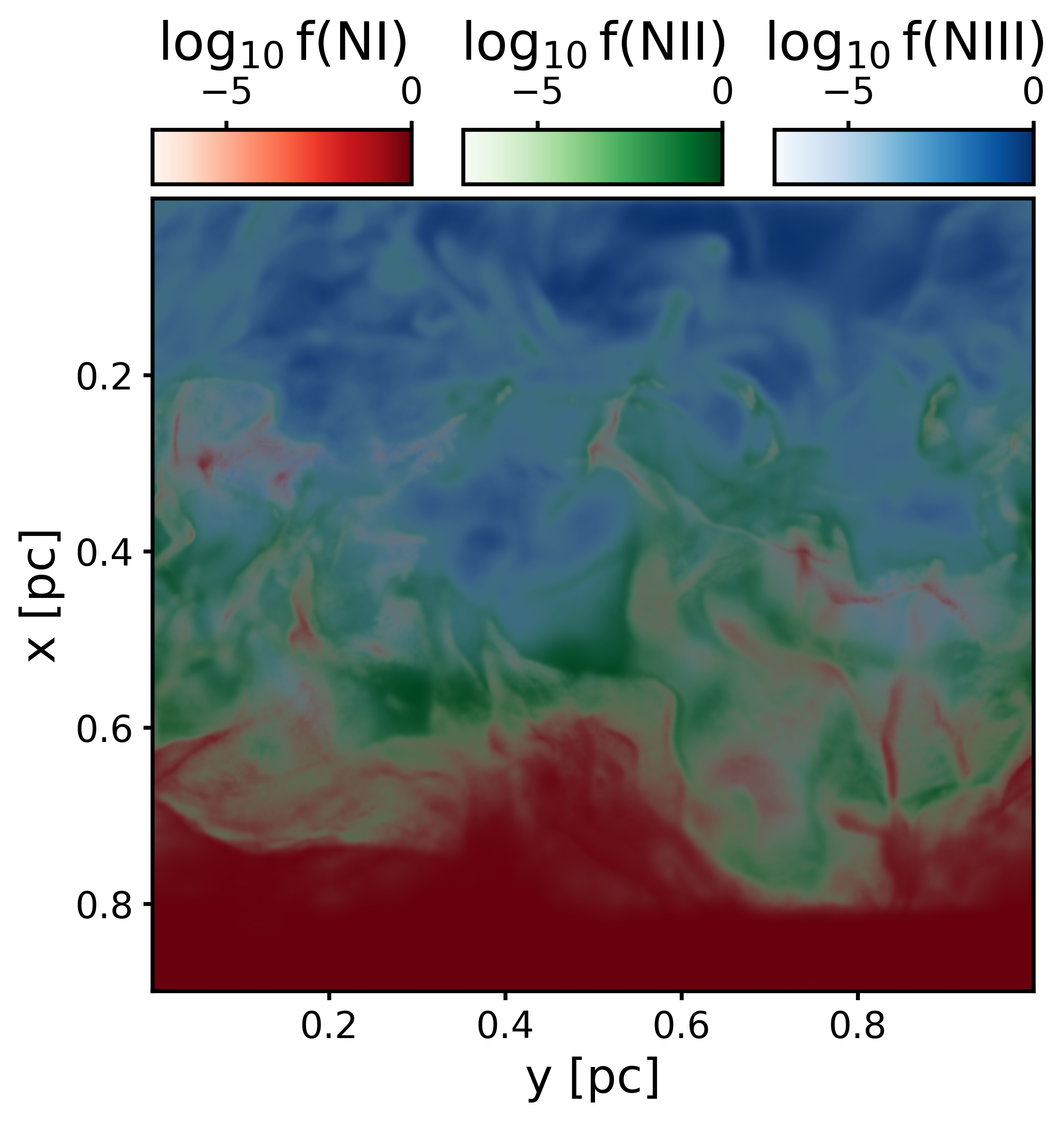}
\vspace{-3.5mm}
\caption{The projected mass-weighted ionisation fractions of \ion{N}{I} ({\it red}), \ion{N}{II} ({\it green}) and \ion{N}{III} ({\it blue}) for the homogenous case ({\it left panel}) and the inhomogeneous case ({\it right panel}), both at time $t\approx 400$~yr since radiation is injected. As in Fig.~\ref{fig:RadBound_rhoT},
radiation propagates along the $x$-axis from the top to bottom, we project along the $z$-axis, and we cut off the areas with $x > 0.9$~pc. Colours scale logarithmically with the ionisation fractions. The non-uniform density distribution results in much more \ion{N}{I} and \ion{N}{II} due to inhomogeneities and shielding.}
\label{fig:RadBound_CombineN}
\end{figure*}

We plot projected density and temperature maps for both simulations at $t=200$~yr in Fig.~\ref{fig:RadBound_rhoT}. In the homogeneous case (left panel), the ionisation front has propagated through approximately 3/4 of the simulation volume, and the density distributions before and after passage of the ionisation front look very similar. This is because
the speed of the ionization front ($\gtrsim 1000 ~{\rm km~s}^{-1}$) is much faster than the hydrodynamics response (which is of the order of the sound speed, $\sim 10~{\rm km~s}^{-1}$) - therefore the gas remains approximately static as it is overrun by the ionisation front. The only small deviations from constant density are due to the finite resolution of the glass-like initial condition. The gas behind the photo-ionization front rapidly rises to the photo-ionization equilibrium temperature, $T\lesssim 10^{4}~{\rm K}$. 

The right panel of Fig.~\ref{fig:RadBound_rhoT} shows the case with the inhomogeneous initial condition. The ionisation front has also propagated through approximately 3/4 of the simulation volume, but the front itself is corrugated due to the underlying density inhomogeneities. These strong density inhomogeneities are apparent in the density field, and again are very similar before and after the passage of the ionisation front, because the gas remains approximately static as it is overrun by the ionisation front. 

The non-uniform densities result in an inhomogeneous temperature distribution, as seen in the right subplot. The low-density gas is rapidly photo-heated to photo-ionization equilibrium with $T\lesssim 10^{4}\;{\rm K}$. In contrast, higher-density gas can self-shield, 
and remain at much lower temperatures, $T\sim 10-10^3~{\rm K}$.

\begin{figure*}
\includegraphics[width=0.98\textwidth]{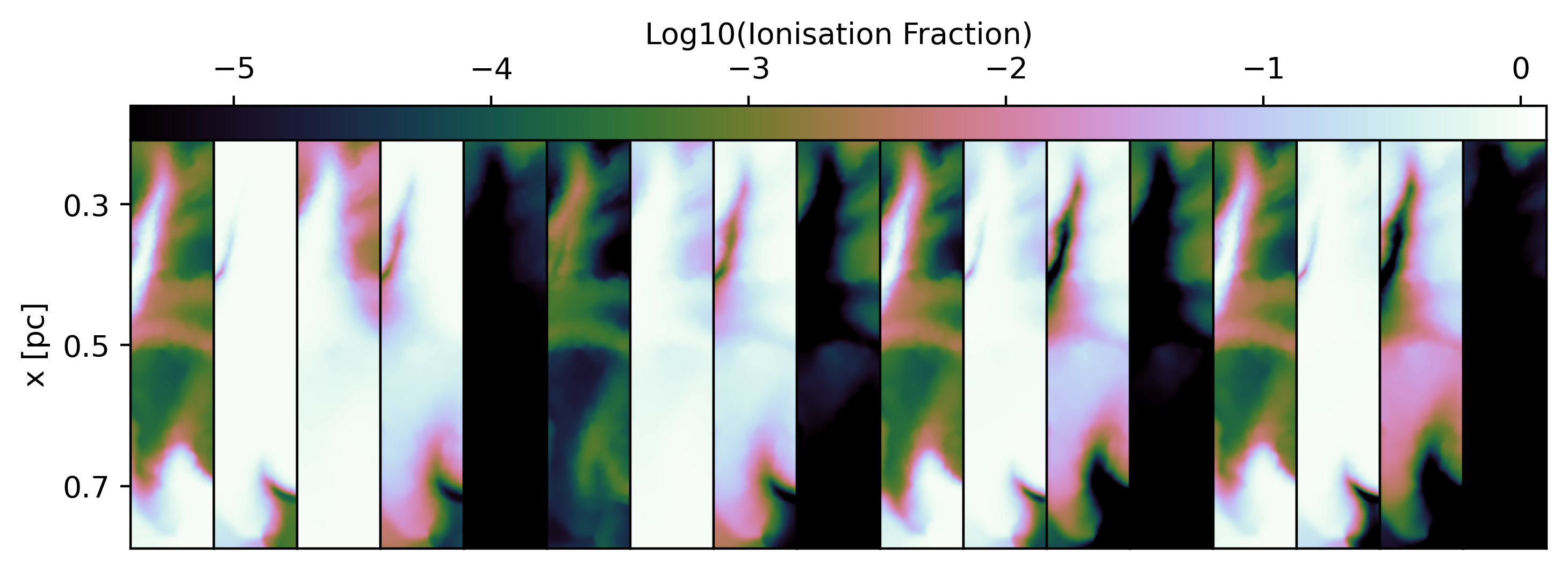}
\vspace{-3.5mm}
\includegraphics[width=0.98\textwidth]{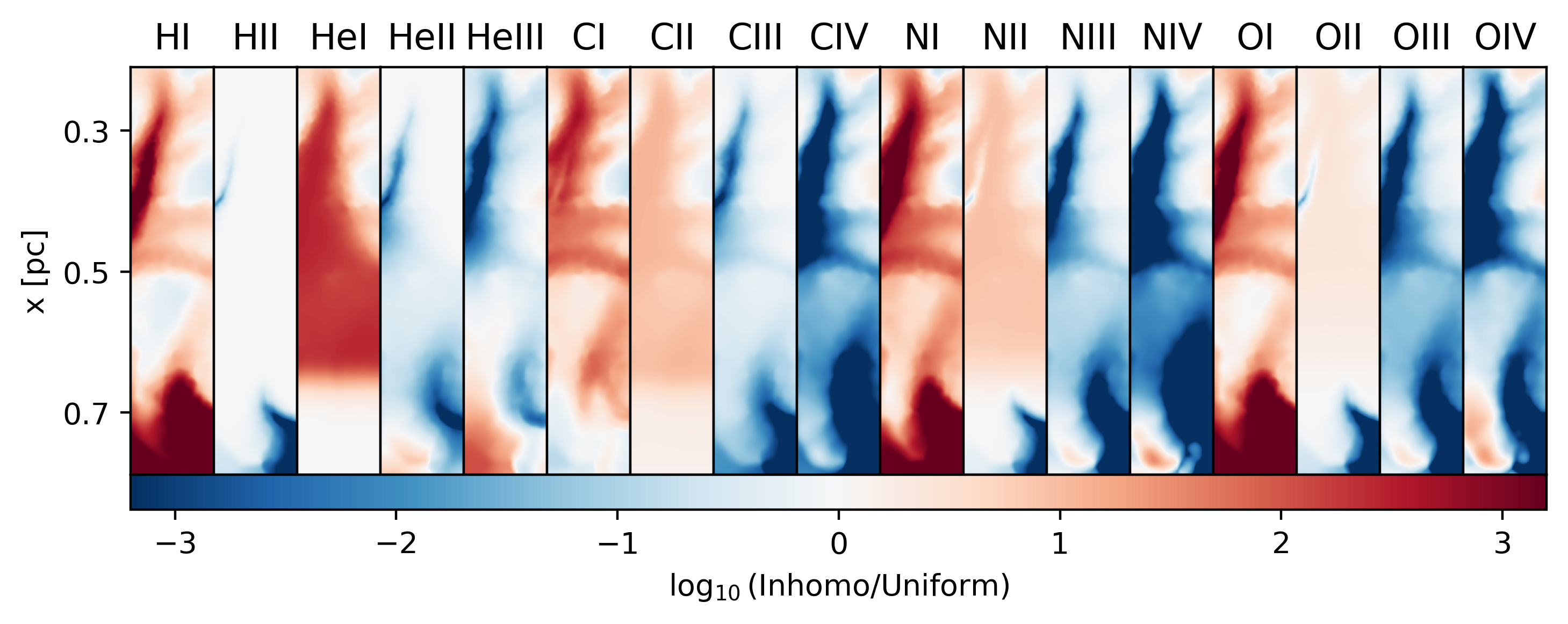}
\caption{ Propagation of an ionisation front through the {\sc ism}. As in Fig.~\ref{fig:RadBound_rhoT}, radiation propagates along the $x$-axis from the top to bottom. {\it Upper panels:} slice-plots of the mass-weighted ionisation fractions from the inhomogeneous run (Inhomo) through the z mid-plane at time t $\sim$ 400 yr. 
{\it Lower panels:} ratios of the ionisation fractions from the inhomogeneous run (Inhomo) over the homogeneous run (Uniform).
For better visualization, we focus only on a small slice through $z=0.5$~pc with $0.2~{\rm pc} < x < 0.8~{\rm pc}$ and $0.7~{\rm pc} < y < 0.8~{\rm pc}$.}
\label{fig:unipystrom_slice}
\end{figure*}

The ionisation fractions are quite different when comparing the uniform and inhomogeneous cases, as shown in Fig.~\ref{fig:RadBound_CombineN} for the case of nitrogen ions at time $t\approx 400$~yr. In both runs, the higher ionisation stages, such as \ion{N}{III}, dominate
close to the injection region, whereas the lower ionisation stages, such as \ion{N}{I}, dominate at $x\ge 0.8$~pc, where the radiation field is much weaker. However, the inhomogeneous run contains significant fractions of low ionisation states (\ion{N}{I} and \ion{N}{II}) surrounded by the higher ionisation stages.
These filaments and sheets of low ions survive the passage of the front due to self-shielding (compare the structures to the high-density region in Fig. \ref{fig:RadBound_rhoT}). 

The fractions of other ions in the inhomogeneous run are shown in the top panel of Fig.~\ref{fig:unipystrom_slice}. Based on these plots,
species can be approximately divided into three groups: low 
ionisation states (\ion{H}{I}, \ion{N}{I}, \ion{O}{I}), intermediate ionisation states (\ion{He}{II}, \ion{C}{III}, \ion{N}{III}), and higher ionisation states (\ion{He}{III}, \ion{C}{IV}, \ion{O}{IV}, higher ions). Within each group, species have similar fractions throughout the volume, since these species have similar ionisation energies. Initially, all elements are mostly in the low ionisation state. They are then rapidly ionised into the intermediate ionisation states by the photons in the lowest EUV radiation bin, but remain neutral in high-density self-shielded regions. The simulation has negligible fractions of ions in even higher ionization states, e.g. \ion{He}{III}, \ion{C}{IV} and \ion{O}{IV}, since the ionising spectrum - a $T=3\times 10^4~{\rm K}$ black body - contains too few high-energy photons to produce these higher states.

We contrast the homogeneous and inhomogeneous cases by
plotting the ratio between ion species fractions in the uniform and the inhomogeneous runs in the lower
panel of Fig.~\ref{fig:unipystrom_slice}. The lower ionization states (e.g. \ion{H}{I}, \ion{He}{I}, \ion{N}{I}) are strongly boosted by inhomogeneities by up to a factor of a thousand, especially in the high-density regions, due to shielding. Consequently, higher ionization states (e.g. \ion{H}{II}, \ion{He}{II}, \ion{N}{III}) are suppressed (up to a factor of a thousand), especially in regions where density is high and shielding is effective. The intermediate ions (e.g. \ion{O}{II}, \ion{N}{II}) have both enhancements at low-density regions and suppression at high-density regions, compared to the homogeneous case. This is consistent with \cite{Jin23spatialHII}, who found the emission lines from \ion{N}{II}, \ion{S}{II}, and \ion{O}{I} are enhanced with inhomogeneities in HII regions, while  \ion{O}{III} emission is suppressed.

\begin{figure}
\includegraphics[width=0.49\textwidth]{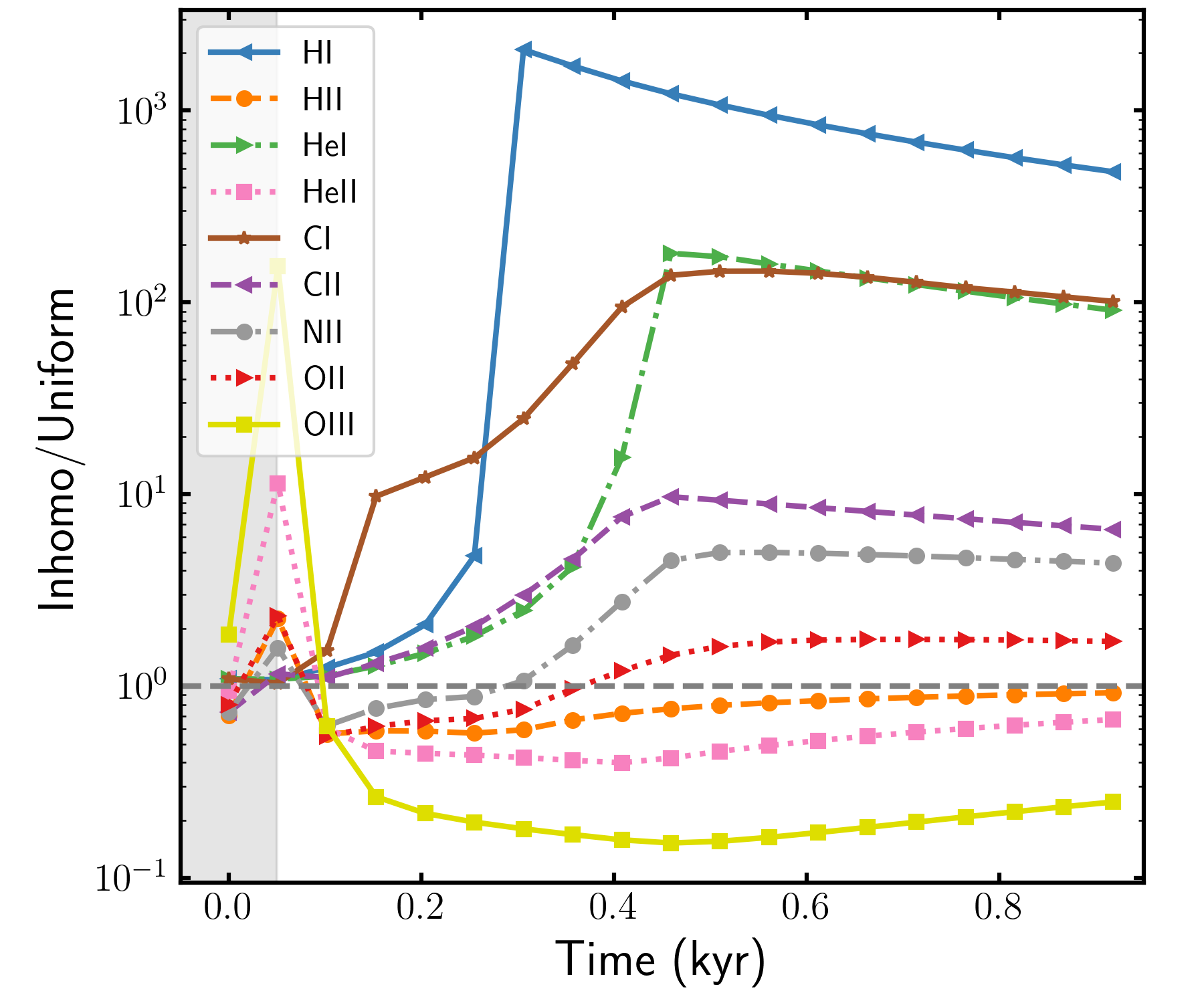}
\vspace{-7mm}
\caption{Ratios of the mean ionisation fractions in the inhomogeneous run over the homogeneous run as a function of time since the radiation is injected. The horizontal grey dashed line indicates where the ratio is unity. Radiation propagates along the $x$-axis and the mean is computed in the region $0.3~{\rm pc} < x < 0.7~{\rm pc}$, to avoid the effects of absorption and radiation injection. The shaded region indicates where the data is unreliable since only a small fraction of the gas is ionized. Inhomogeneities and self-shielding can create differences of several orders of magnitude between the ionisation fractions in the two cases.}
\label{fig:RadBound_ratio}
\end{figure}

A more quantitative illustration of the effect of inhomogeneities and self-shielding on ionisation fraction is provided by Fig.~\ref{fig:RadBound_ratio}.
Ionisation fractions change rapidly during the initial fast ionization front propagation before $t\sim 0.4~{\rm kyr}$. At later times, the ionization fronts slowly penetrate and photo-evaporate dense gas, so some dense structures will be ionized and the ionisation fractions get closer to those in the uniform case. However, especially thanks to self-gravity, some of the over-dense regions are likely to fragment and accumulate more mass\footnote{ Indeed, ionisation fronts could even boost over-densities, induce gravitational collapse, and trigger star formation \citep{Bisb11RDI,Grit09turbionisation}.}.

The ratios follow similar trends as seen in Fig.~\ref{fig:unipystrom_slice}: inhomogeneities boost low ions (e.g. \ion{H}{I}) and suppress higher ions (\ion{O}{III}). Overall, this comparison illustrates the importance of including and {\em resolving} inhomogeneities when studying or interpreting ionisation fraction in the interstellar medium.

\subsection{Importance and signatures of non-equilibrium effects}
\label{sec:comequ}

\begin{figure*}
\includegraphics[width=0.98\textwidth]{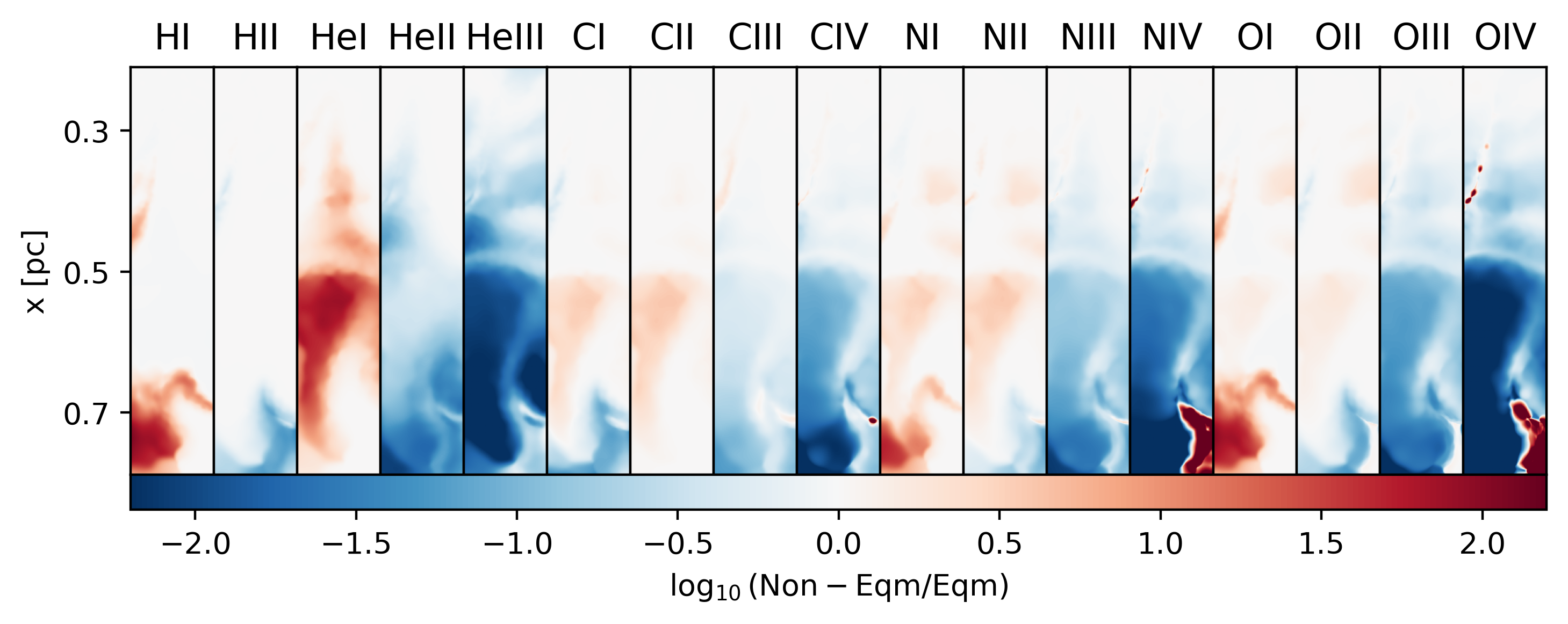}
\caption{ Propagation of an ionisation front through the {\sc ism}. As Fig.~\ref{fig:unipystrom_slice} (lower panel), but plotting the ratios of ionisation fractions for 
the non-equilibrium case (Non-Eqm) over that for the equilibrium case (Eqm) at time $t\approx 400$~yr. In general, there are more low ions (e.g. \ion{H}{I}, \ion{N}{I}) and fewer high ions (e.g. \ion{H}{II}, \ion{N}{III}) in the non-equilibrium case compared to the equilibrium case.}
\label{fig:eqmpystrom_slice}
\end{figure*}

The runs presented so far use \swifter{}'s ability to evolve rate equations, meaning that ionisation fractions can be out of equilibrium. Here we compare these results to the ionisation fractions in \lq chemical equilibrium\rq. There are various ways to define when gas is in \lq equilibrium\rq. In \lq chemical equilibrium \rq, we calculate the equilibrium ionisation fractions using the radiation field, temperature, and density from the non-equilibrium simulations \cite[see also][]{Rich22raddust}. We note that chemical equilibrium is not the same as thermodynamic equilibrium.

In practice, we determine the \lq chemical equilibrium\rq\ ionisation fractions by evolving all gas particles using \chimes{} for 1~Gyr, keeping temperature, density, and the radiation field constant. We verify that the equilibrium values obtained in \swifter{} agree with those obtained
directly from \chimes\ in Appendix \ref{sec:verequcal}.

We contrast the non-equilibrium and equilibrium cases by
plotting the ratio between ion species fractions in slices in Fig. \ref{fig:eqmpystrom_slice}, which is similar to Fig.~\ref{fig:unipystrom_slice}. In general, there are more low ions (e.g. \ion{H}{I}, \ion{He}{I}, \ion{O}{I}) and fewer higher ions (e.g. \ion{He}{III}, \ion{O}{III}, \ion{C}{IV}) in the non-equilibrium case compared to the equilibrium case. This is because it takes some time to photo-ionize these low ions.

We plot ratios of the mean ionisation fraction for the two cases in Fig.~\ref{fig:RadBound_ratio_eqm}, which is similar to Fig.~\ref{fig:RadBound_ratio}. In non-equilibrium runs, ionisation fractions of low ions (\ion{H}{I}, \ion{He}{I}, \ion{O}{I}) are enhanced, while those of higher ions are correspondingly reduced. 
Non-equilibrium effects change the ionisation fraction of some ions by tens of per cent over time-scales of
around kyrs.

\subsection{Mock {\sc ism} observations generated with \YKcode{} and \radmc{}}
\label{sec:radmc3d}

Here we show how simulations performed with \swifter{} can be post-processed to generate forward models that can be directly compared to observations, for example, to {\sc ifu} data.

We compute line emissivities for transitions in ions as follows. For recombination lines such as those of hydrogen, we use the \YKcode{} code \citep{Liu25Hylight} to determine the level population of \ion{H}{I}, given the density, temperature, and ionisation fraction extracted from the \swifter{} run. 
\YKcode{} accounts for level-resolved recombinations, the cascade of the electron from higher to lower levels in the hydrogen atom, as well as collisional excitation.
We interpolate the computed emissivities from the SPH particles to an adaptive mesh refinement ({\sc amr}) grid, and then use \radmc{}, a Monte Carlo radiative transfer code\footnote{\url{https://www.ita.uni-heidelberg.de/~dullemond/software/radmc-3d/}} \citep{Dullemond12}, to compute line luminosities and surface brightness for commonly observed lines, such as H$\alpha$ and H$\beta$.

For other ions, we interpolate the ionic densities from the SPH particles directly to the {\sc amr} grid, and use \radmc{} to compute emissivities in the optically thin non-LTE limit, and from there, luminosities and surface brightness.

Fig.~\ref{fig:RadBound_TurbBox_RGB_Cropped} shows an example mock observation of the inhomogeneous
{\sc ism} simulation, focussing on the surface brightness in the [\ion{N}{II}]$\lambda6584$, ${\rm H}\alpha$ and [\ion{O}{III}]$\lambda5007$ nebular lines. As before, radiation is injected into the volume at the top and streams down along the $x$-axis, and the region is \lq observed\rq\ along the $z$-axis. The surface brightness of [\ion{O}{III}]$\lambda5007$ is high at the top of the region, where the radiation is injected. [\ion{N}{II}] emission arises mostly from gas close to the ionization
front (about 3/4 down from the top), and in filaments of higher density. Performing a more detailed analysis of these types of mock observations, including comparison to observations, is beyond the scope of the present paper.

\begin{figure}
\includegraphics[width=0.49\textwidth]{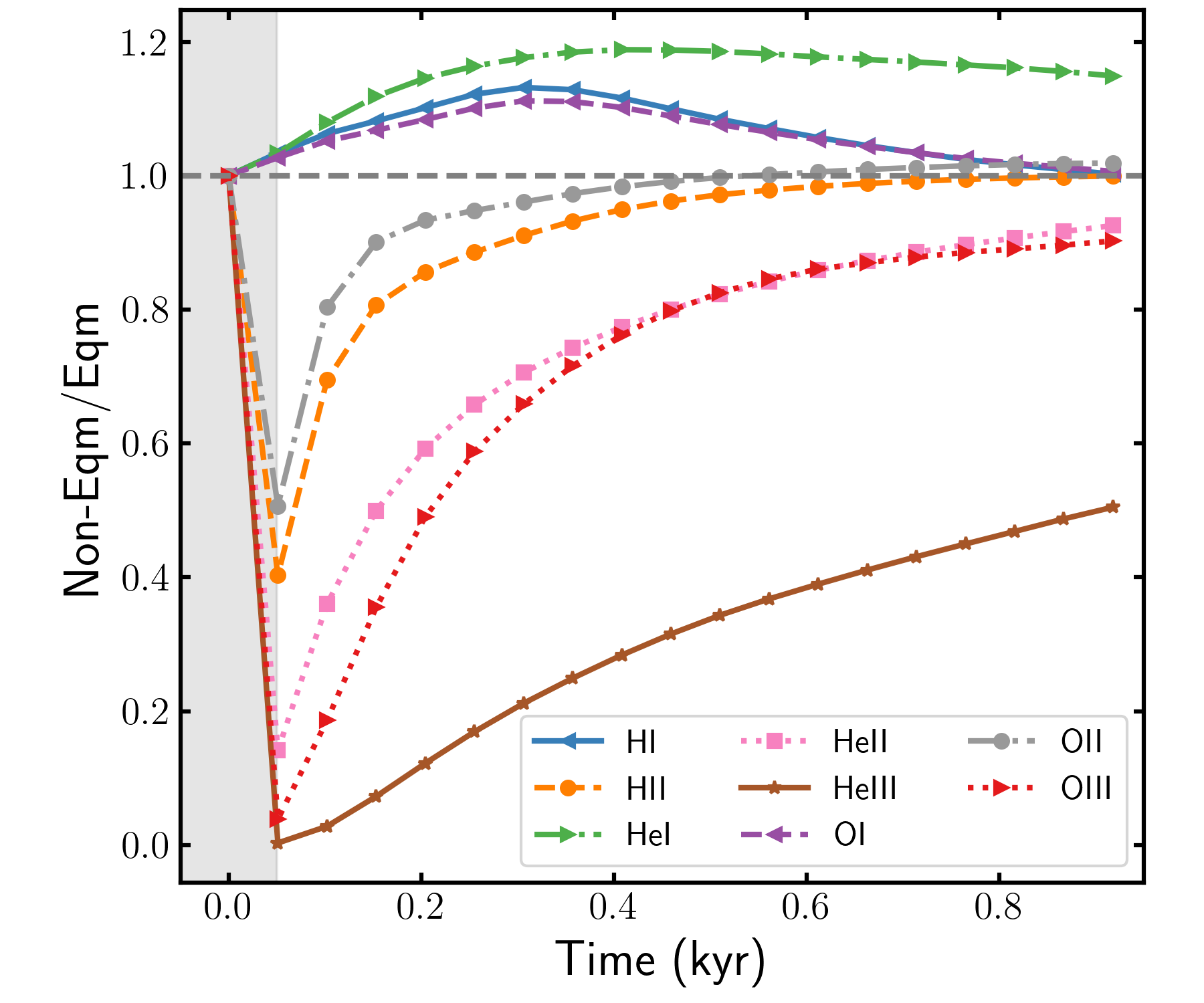}
\caption{As Fig.~\ref{fig:RadBound_ratio}, but plotting the ratios of the mean ionisation fractions in the non-equilibrium case (Non-Eqm) over the equilibrium case (Eqm). Non-equilibrium effects change some ionisation fractions by up to 10s of per cent. As time progresses, the non-equilibrium fractions approach the equilibrium fractions for all ions.}
\label{fig:RadBound_ratio_eqm}
\end{figure}

\begin{figure}
\includegraphics[width=0.49\textwidth]{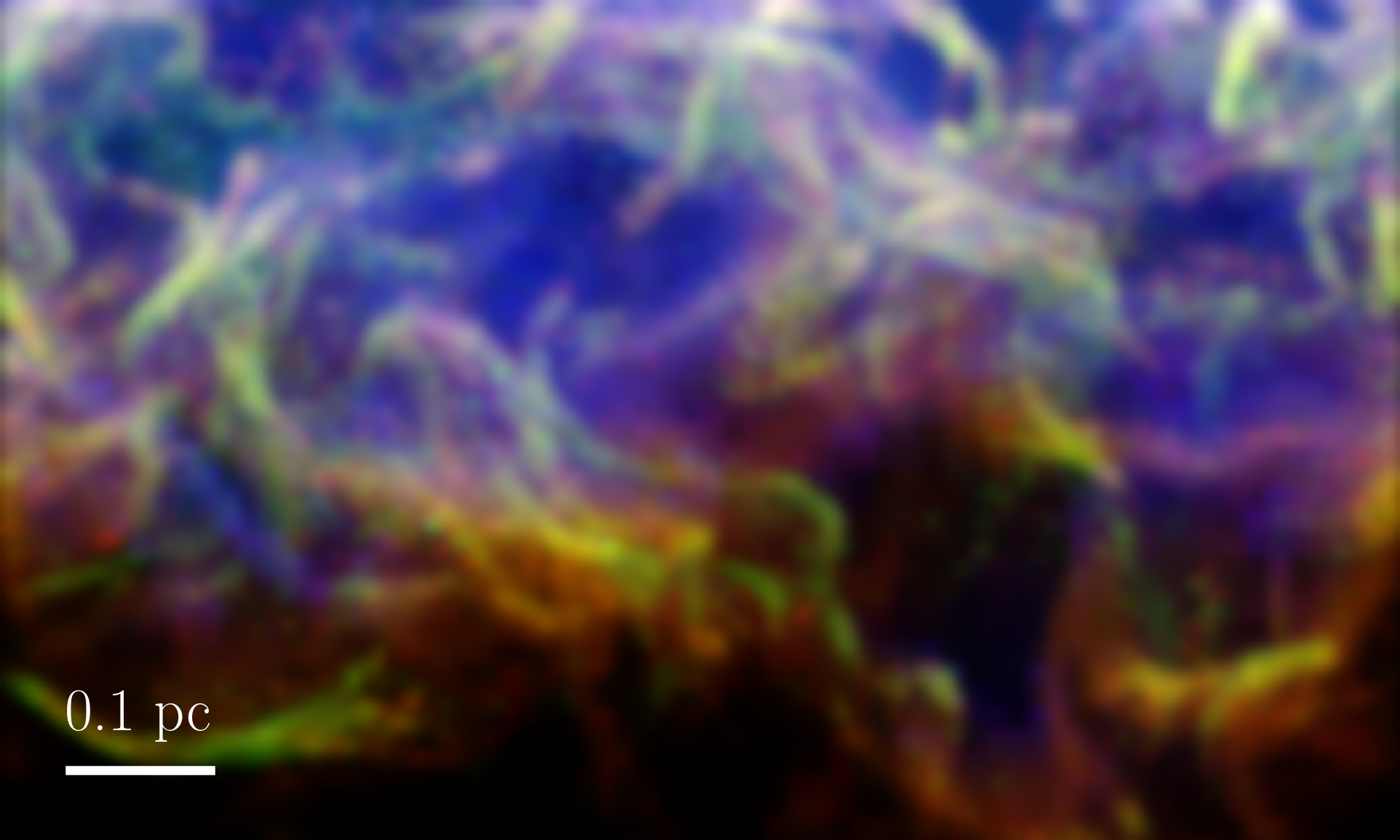}
\caption{Mock observation image of our inhomogeneous interstellar medium, generated with \radmc{} at time $t\sim 0.5\;{\rm kyr}$. Red, green, and blue colors represent [\ion{N}{II}]$\lambda$6584, ${\rm H}\alpha$, and [\ion{O}{III}]$\lambda$5007 respectively. We only plot regions from x = 0.2 pc (top) to 0.8 pc (bottom), in order to crop out the injection and absorption regions in this image. We applied a Gaussian filter with a smoothing length of 0.02 pc to simulate the effect of a point spread function.
}
\label{fig:RadBound_TurbBox_RGB_Cropped}
\end{figure}

\section{Timing benchmarks}
\label{sec:timing}

\begin{figure}
\includegraphics[width=0.48\textwidth]{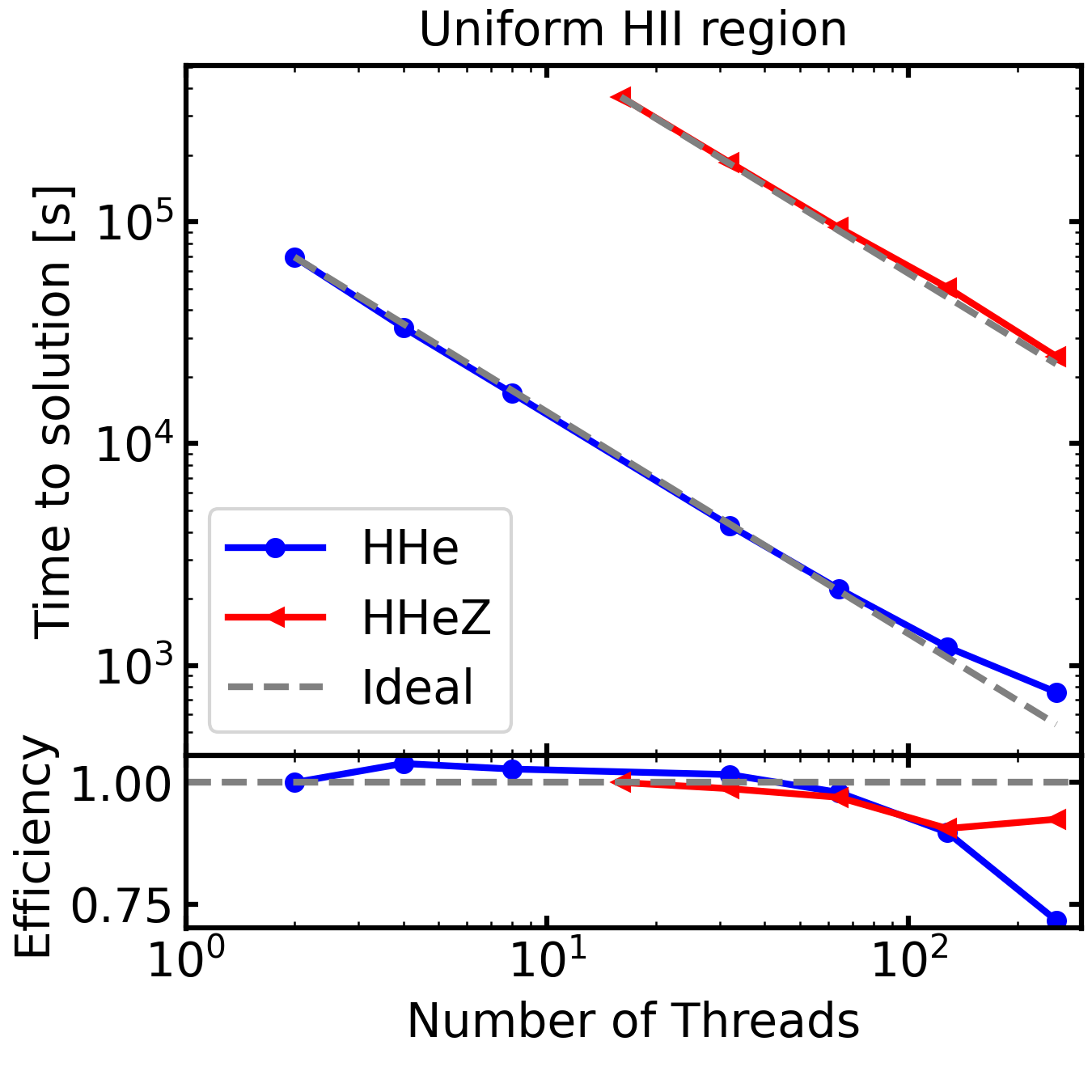}
\includegraphics[width=0.48\textwidth]{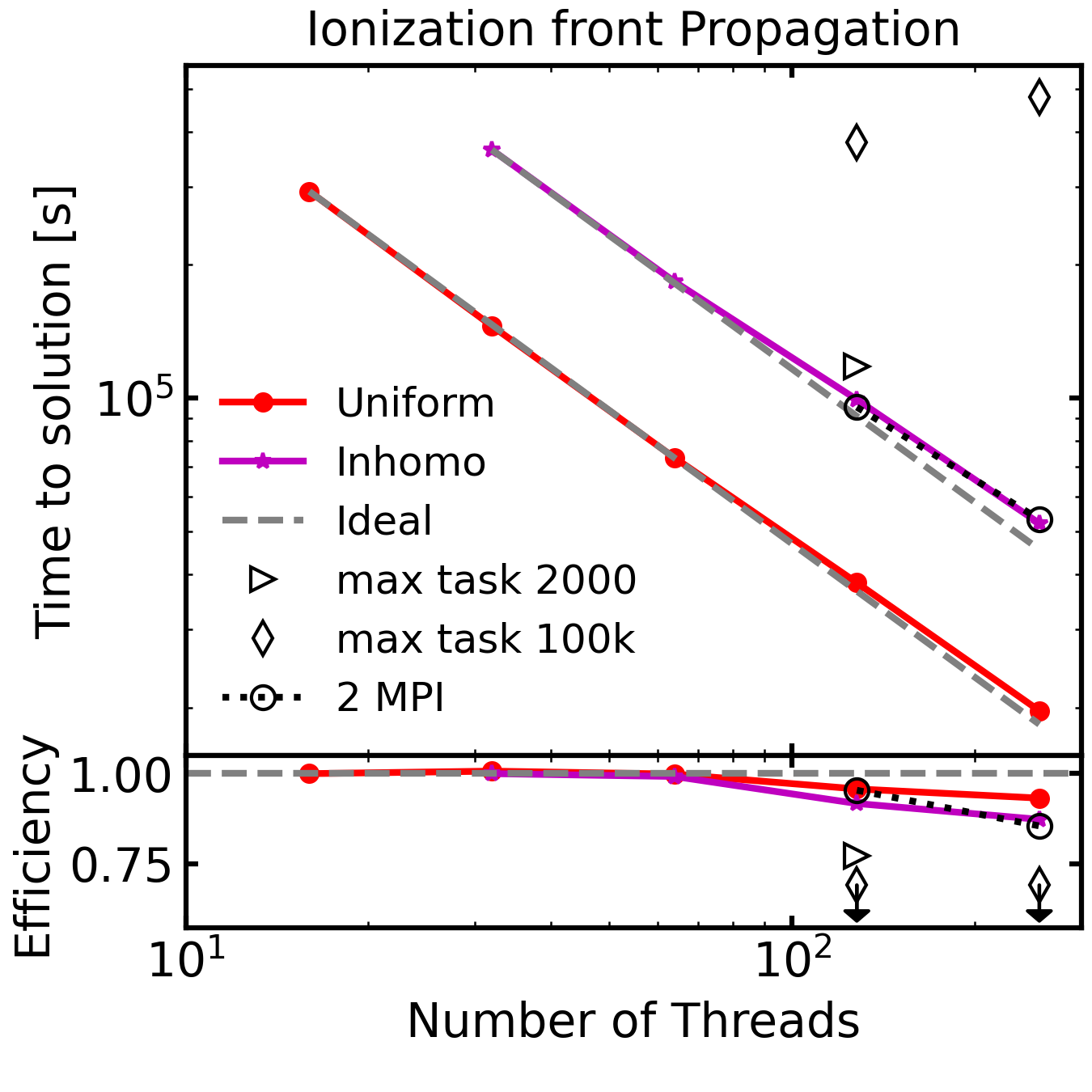}
\caption{Strong scaling tests. The {\it top panel} shows the multi-frequency \ion{H}{II} region ($128^3$ particles) with (HHe; blue circle lines) and without metals (HHeZ; red triangle lines). The {\it bottom panel} shows the propagation of the ionization ($128^3$ particles) through the uniform (red circle lines) and inhomogeneous (purple star lines) interstellar medium (see the setup in \S\ref{sec:application}). The grey dashed lines represent the ideal results if the code can scale perfectly (based on the run time of the fewest-thread simulations).  The empty diamond and triangle symbols represent simulations with a higher maximum number of particles per task (see the main text). The empty circle dotted lines represent inhomogeneous runs with two {\small MPI} but the same total number of threads. The downward arrows indicate that the data is under the y-axis lower limit. {\bf For both panels}: {\it Top sub-panels:} the time to solution in seconds. {\it Bottom sub-panels:} Parallel Efficiency, as defined as the ratio between the simulation time and the run time assuming the ideal scaling. }
\label{fig:strongscaling}
\end{figure}

In this section, we examine the computational performance of \swifter{}, including the parallel efficiency over many CPU cores, using {\small POSIX} threads or in combination with {\small MPI}. We examine two sets of problems: (1) the \ion{H}{II} regions in a uniform medium and (2) ionization front propagation in the inhomogeneous/uniform interstellar medium with self-gravity and hydrodynamics enabled. The first set also examines the effect of metals on computational times, whereas the second set tests the code performance in an inhomogeneous medium.

In the upper panel of Fig. \ref{fig:strongscaling}, we consider the multi-frequency \ion{H}{II} region test similar to the test in \S\ref{sec:HII}. Unlike \S\ref{sec:HII}, the simulation box here contains $128^3$ gas particles, and we run the simulations for 5 Myr. We also consider another run with the same setup but with element abundances described in Table \ref{table:eleab}. All of the tests are run on {\small AMD} {\small EPYC} 9754 CPUs\footnote{Our {\small AMD} {\small EPYC} is a high-performance server processor with 128 cores per socket and 2 sockets per node.}.

We perform a strong scaling test by running the simulations with $128^3$ particles with different numbers of threads (from 2 to 256 threads; with one thread per core). The performance is shown in the upper panel in Fig. \ref{fig:strongscaling}. Compared to the hydrogen+helium (HHe) runs, the full metal run (HHeZ) takes approximately 50 times longer, since the network size increases from 10 to 157. The parallel efficiency is higher than 90\% up to 100 cores for two million particles without metals. With metals, the parallel efficiency is even higher due to more work to be done. 

In the lower panel in Fig. \ref{fig:strongscaling}, we test the strong scaling performance for the ionization front propagation. The setup is identical to \S\ref{sec:application}, e.g., with the same physics (i.e., hydrodynamics and self-gravity), resolution, initial condition, and radiation flux. 

Notice that the inhomogeneous run has a deep time-step hierarchy. Over 50\% of particles have shorter time-steps compared to the uniform run, while a small fraction of particles have longer ones. 20\% particles has 1/4 shorter time-steps, whereas 1\% particles has 1/8 shorter. The time-step range spans a factor of $\sim 60$. This hierarchy poses a challenge to efficiently distribute the computational work across many CPU cores. 

To parallelize the work efficiently, we applied the fine-grained tasking techniques described in \S\ref{sec:swiftimplement}, with the maximum number of particles per RT thermo-chemistry task $=50$. However, we only run the simulations for 0.285 kyr when the ionization front goes through approximately half of the box. The inhomogeneous (inhomo) runs take approximately four times longer than the uniform runs, since extra steps are required to evolve the denser particles. 

Both the uniform and inhomogeneous runs have excellent scaling efficiency ($\sim$85-90\%; for 8000 particles per thread)\footnote{ For reference, one CPU core can update around 18,000 particles per second without metals, whereas only 400 particles per second in the full metal run, i.e., the full metal run is $\sim 50$ times more expensive. Notice that these runs include hydrodynamics, self-gravity, radiative transfer, and thermo-chemistry network.}. The scaling performance is consistent with the \swift\ strong scaling behavior in \cite{Scha16SWIFT}. This high parallel efficiency is due to our fine-grained tasking techniques (\S\ref{sec:swiftimplement}). In particular, we split the RT thermo-chemistry into smaller units if the number of particles per task exceeds 50, which is crucial to the parallelization. We demonstrate this by running simulations with larger maximum numbers of particles per task, as represented by empty symbols in Fig. \ref{fig:strongscaling}. ``max task 2000'' and ``max task 100k'' represent 2000 and 100,000 particles, respectively. The run time can be an order of magnitude longer if we do not split the RT thermo-chemistry into smaller units (in the 256-thread case).

Our code is also capable of {\small MPI} parallelization\footnote{Notice that we still run on a single node but across multiple sockets. The optimization of inter-node communication of our module on the CUHK central cluster will be addressed in future work.}. To demonstrate this, we run the inhomogeneous runs with 2 {\small MPI}, each with 64 or 128 threads. Their run times are plotted as the empty circle dotted lines in the lower panel of Fig. \ref{fig:strongscaling}. The parallelization performance is excellent and comparable to pure thread-parallelization runs. 

We have checked that most of the run time (>90\%) is spent on solving the thermo-chemistry equation with the full metal network. As a consequence, sub-cycling of radiative transfer (\S\ref{sec:swiftimplement}) does not significantly reduce the compute time.

\section{Summary and Outlook}
\label{sec:conclusion}
In this work we described and tested the \swifter{} code, which couples the \chimes{} non-equilibrium thermo-chemistry network (described in \S\ref{sec:CHIMES_method}, \citealt{Rich14CHIMESI}) with the radiative transfer method \swiftrt{} (described in \S\ref{sec:SPHM1RT}, \citealt{Chan21SPHM1RT}), and
the \swift{}  cosmological hydrodynamics code \citep{Scha23SWIFT}. The implementation ({\em i}) couples radiation hydrodynamics with non-equilibrium thermo-chemistry with 11 elements that are important in the {\sc ism} and {\sc cgm} of galaxies, all their positive ions and some negative ions as well (\S\ref{sec:chemistryeq}), ({\em ii}) handles multi-frequency radiation using an arbitrary number of radiation bins, ({\em iii}) accounts for dust attenuation, grain recombination, photoelectric heating by grains, and cosmic ray ionisations (\S\ref{sec:dustmethod}), and ({\em iv}) calculates direct radiation pressure on both gas and dust (\S\ref{sec:dirradpre}).

We verified \swifter{} using a series of tests: 
\begin{enumerate}
    \item In \S\ref{sec:iliev0}, we presented a 
    thermo-chemistry test for a single parcel of hydrogen gas, and compared the results from \swifter{}
    against analytic solutions. We verified that the multi-frequency thermo-chemistry solver can correctly capture the non-equilibrium evolution of the heating and ionisation of hydrogen gas irradiated by a fixed radiation field.

    \item We tested the direct radiation pressure implementation in \S\ref{sec:radpretest}. Our simulated ionized gas velocities driven by radiation pressure match well with the \cite{Wise12radpre} analytic solutions.

    \item In \S\ref{sec:Stromgren}, we presented a test of an \ion{H}{II} region consisting of hydrogen and helium only and of an \ion{H}{II} region where the gas has solar abundances of elements. Once ionisation fractions are in equilibrium, in the solar abundance test, the results from \swifter{} agree well with those from {\sc cloudy} (version c22.02,  \citealt{Ferl17cloudy}), which assumes equilibrium. When the ionisation fractions are
    not in equilibrium yet, \swifter{} agrees well
    with the results of the {\sc tt1d} spherically symmetric non-equilibrium multi-frequency photo-ionization code \citep{Pawl08TRAPHIC} in the hydrogen and helium only test.

\end{enumerate}

In \S\ref{sec:application}, as an example application of \swifter{}, we studied the propagation of an ionisation front through an {\sc ism}-like gas distribution, comparing the case of a homogeneous {\sc ism} to an inhomogeneous {\sc ism}, and comparing a non-equilibrium simulation against an equilibrium calculation. Inhomogeneities can boost the low ions (e.g. \ion{H}{I}) and suppress the higher ions (e.g. \ion{C}{III}) by factors of a thousand. Non-equilibrium effects can boost the abundance of low ions and reduce that of higher ions by tens of percent, but these effects decrease in time (\S\ref{sec:comequ}).

We generated mock images of the inhomogeneous {\sc ism} calculation in the [\ion{N}{II}]$\lambda$6584, ${\rm H}\alpha$, and [\ion{O}{III}]$\lambda$5007 nebular lines (\S\ref{sec:radmc3d}). 
We use \YKcode{} \citep{Liu25Hylight} to compute level population for the \ion{H}{I} atom and predict the H$\alpha$ emissivities. Emissivities for the other ions, and mock images, are generated with the 3D \radmc{} Monte Carlo radiative transfer code \citep{Dullemond12}.

Finally, we showed in \S\ref{sec:timing} that \swifter{} has excellent parallel efficiency (strong scaling). With $128^3$ SPH particles with all available metal elements, our code maintains 90\% parallel efficiency for over 200 cores in our highly inhomogeneous interstellar medium simulations.

We now outline current limitations of \swifter{} and potential avenues for future work.

\begin{enumerate}[label=(\Alph*)]
\item We neglected molecules when coupling \chimes{} to radiative transfer in \swifter{}, even though the \chimes{} network includes them.
To enable a successful integration of the \chimes{} molecular network within \swifter{} required  
tracking the propagation of the Lyman-Werner (photon energies between $11.2\;{\rm eV}<h\nu<13.6\;{\rm eV}$) ${\rm H}_2$ dissociating photons. The molecular bands associated with ${\rm H}_2$ make it challenging to account for self-shielding. Often, self-shielding is approximated with the analytic fitting function from \cite{Drai96H2shield}. Unfortunately, the ${\rm H}_2$ column density - which enters the self-shielding fitting function - is not (yet) available during the simulations. One option is to approximate that column density using a Sobolev-like length scale, as suggested by \cite{Gned09H2shield}. Another possibility is to follow the suggestion of \cite{Nick18H2RT} to mimic self-shielding by enhancing the destruction of the Lyman-Werner photons.

\item The current treatment of dust is very simplified: we assume a constant dust-to-gas ratio that depends on metallicity, rather than accounting for the formation and destruction of dust grains. A significant amount of metals is locked in dust grains \citep[e.g.][] {Jenk09dustdepletion} and dust absorbs UV light and radiates in the IR. Accounting for both would improve the realism of {\sc ism} simulations performed with \swifter{}.

A natural extension of the model would involve incorporating a \lq live\rq\ dust models \citep[e.g.][]{McKi16dustI,Chob22dust}), which accounts for the production of dust by stars, metal depletion onto dust, destruction of dust in supernovae, thermal sputtering in hot gas, etc. For example, \cite{Tray25dustISM} coupled a live dust model with \chimes{} in \swift{}. Alternatively, \cite{Rich22raddust} developed an empirical density-based dust model using observed dust depletion factors, one step beyond assuming a constant dust-to-metal ratio.

\item The sources of radiation are currently restricted to simple black body spectra that do not evolve. One way to improve upon this is to use the comprehensive grid of O-type star spectra from \cite{Lanz03OSTAR}. Alternatively, \swifter{} could be interfaced with the stellar evolution tracks and stellar atmospheres extracted from {\small STARBURST} \citep{Leit99starburst99} or {\small BPASS} \citep{Byrn22BPASS}.

\item \swifter{} currently follows the interaction of radiation with the gas through photo-ionisation, photo-heating, and radiation pressure. However, realistic simulations of the {\sc ism} should also include the impact of stellar winds, supernovae, and magnetic fields (amongst others). We aim to implement at least some of these in the near future.

\item The computational efficiency of the radiative transfer module, and in particular that of the \chimes{} network, could be improved further. In \S\ref{sec:timing}, we demonstrate that \swifter{} parallelizes well in the uniform \ion{H}{II} region and ionization front propagation through an inhomogeneous medium. However, the full \chimes{} network is much slower than the network with only hydrogen and helium. 
\chimes{} is currently being developed to utilise accelerated thermo-chemistry solvers, for example, using GPUs (e.g. \citealt{Balo21GPUSUNDIALS}) or using less computationally intensive ODE integration methods (e.g. \citealt{Anni97multchem,Katz22RTZ}). The variable speed of light approximation can be another possibility to accelerate our simulations \citep{Chan23mhreion}.

\end{enumerate}

There are several promising applications of \swifter{} in astronomy. Expanding on the example of an ionization front propagating through the {\sc ism} (\S\ref{sec:application}), we plan to explore the parameter space, examining the effects of different source spectra, geometries, and density distributions on temperatures and ionisation fractions. We envision studying nebular line emission from the {\sc ism} and from \ion{H}{II} regions in terms of line ratio diagrams (such as the BPT diagram of \cite{Baldwin81}). Other possible applications include the study of shocked flows \citep{Teci08NEQ} or Wolf-Rayet nebulae \citep{Math24noneq}. 
Other promising avenues include the study of the impact of radiation and non-equilibrium effects on the circumgalactic medium ({\sc cgm}) of galaxies, as in e.g. \cite{Oppe18AGN}, as well as the observational diagnostics of outflows from active galactic nuclei \citep[e.g.][]{Rich18AGNH2}. Finally, it is also possible to confine the \chimes{} network to primordial species to explore star formation in the early universe and the reionisation of the universe.

\section*{DATA AVAILABILITY}
The \chimes{} code and the associated package are publicly available at \url{https://richings.bitbucket.io/chimes/home.html}. The public version of \swift{} code is available at \url{https://swiftsim.com/}, which includes the \swiftrt{} radiative transfer module coupled with a simple hydrogen and helium thermo-chemistry network. \YKcode{} can be found at \url{https://github.com/YuankangLiu/HyLight}. Currently, \swifter{} is a \swift{} module located in a private \swift{} branch. We plan to release \swifter{} to the public in the future. The data underlying this article will be shared on reasonable request to the corresponding author (TKC).

\section*{ACKNOWLEDGMENTS}
We thank Andrey Kravtsov, Nick Gnedin, Harley Katz, and Anna McLeod for their helpful discussions.

TKC is supported by the `Improvement on Competitiveness in Hiring New Faculties' Funding Scheme (4937210-4937211-4937212), the Direct Grant project (4053662,4443786,4053719) from the Chinese University of Hong Kong, and the RGC Early Career Scheme (24301524). TKC was supported by the E. Margaret Burbidge Prize Postdoctoral Fellowship from the Brinson Foundation at the Departments of Astronomy and Astrophysics at the University of Chicago. 

We acknowledge the support of the CUHK Central High-Performance Computing Cluster, on which the computation in this work has been performed, and we thank Edward So, Anny Cheung, and Nicky Leung for their assistance.

This work also used the DiRAC@Durham facility managed by the Institute for Computational Cosmology on behalf of the STFC DiRAC HPC Facility (www.dirac.ac.uk). The equipment was funded by BEIS capital funding via STFC capital grants ST/K00042X/1, ST/P002293/1, ST/R002371/1 and ST/S002502/1, Durham University, and STFC operations grant ST/R000832/1. DiRAC is part of the National e-Infrastructure. 

This work is co-funded by the European Union (Widening Participation, ExGal-Twin, GA 101158446). Views and opinions expressed are, however, those of the author(s) only and do not necessarily reflect those of the European Union. Neither the European Union nor the granting authority can be held responsible for them. This work received funding from the Horizon Europe guarantee scheme of UK Research and Innovation (UKRI).

The research in this paper made use of the \swift{} open-source simulation code
(http://www.swiftsim.com, \citealt{Scha18SWIFTascl}) version 1.0.0.
This work also made use of matplotlib \citep{Hunt07matplotlib}, numpy \citep{vand11numpy}, scipy \citep{Jone01scipy,Virt20scipy}, swiftsimio \citep{Borr20swiftsimio}, sphviewer \citep{Beni15SPHviewer}, and pyCloudy \citep{Mori13pyCloudy}, the ArXiv repository and NASA’s Astrophysics Data System.

\bibliographystyle{mnras}
\bibliography{mn-jour,mybib}

\appendix

\section{Issues with the Reduced Speed of Light Approximation}
\label{sec:issueRSL}

The reduced speed of light approximation (RSL) is to replace the physical speed of light in the radiative transfer equations with a ``reduced'' value (see \S\ref{sec:rsl}). It is a surprisingly robust approximation, especially when the dynamical time scale is slower than the reduced speed of light, given that the radiation field is slowly varying.

However, we found that RSL cannot capture gas cooling and recombining with a rapidly dimming radiation source. 
We consider a simplified case (\S\ref{sec:test0off}): a hydrogen gas parcel is initially in equilibrium under a radiation source, but the source is turned off at $t=t_{\rm turn-off}$. This process is governed approximately by:
\begin{align}
\frac{\partial \tilde{n}_{\gamma}}{\partial t} \approx&-\sigma_{\rm HI} \tilde{c} \tilde{n}_{\gamma} n_{\rm HI}.
\label{eq:test0offtilde}
\end{align}
The photon density evolution is approximately $\tilde{n}_{\gamma}\propto \exp(-\sigma_{\rm HI} \tilde{c} n_{\rm HI})$, so the decay time-scale of photon is $\sim (\sigma_{\rm HI} \tilde{c} n_{\rm HI})^{-1}$. Reducing the speed of light ($\tilde{c}$) leads to an inaccurate decay time-scale much longer than the correct value $\sim (\sigma_{\rm HI} c n_{\rm HI})^{-1}$. This argument also implies that RSL fails if the radiation source has a rapidly decreasing or varying radiation injection rate. 

We verify this argument by repeating the Single Gas Parcel Thermo-chemistry test with different reduced speed of light $\tilde{c}$, especially after the source is off (\S\ref{sec:test0off}), using our numerical code. We consider pure hydrogen gas and the on-the-spot approximation. Fig. \ref{fig:issueRSL} shows it takes a much longer time to return to neutral hydrogen with a lower $\tilde{c}$. Only the test with the physical speed of light $\tilde{c}=c$ can match the analytic solution (by solving Eq.~\ref{eq:test0offtilde} analytically with $\tilde{c}=c$). 

This issue does not arise with a constant or increasing radiation injection rate, which is governed by: 
\begin{align}
\frac{\partial \tilde{n}_{\gamma}}{\partial t} \approx&-\sigma_{\rm HI} \tilde{c} \tilde{n}_{\gamma} n_{\rm HI}+S_\gamma,\nonumber\\
\frac{\partial n_{\rm HI}}{\partial t} \approx&-\sigma_{\rm HI} \tilde{c} \tilde{n}_{\gamma} n_{\rm HI}+\alpha_{\rm B}n_{\rm HII}^2.
\label{eq:test0ontilde}
\end{align}
The characteristic time scale (to reach photo-ionization equilibrium) is $\sim (\sigma_{\rm HI} \tilde{c} \tilde{n}_{\gamma})^{-1}$, which is independent of $\tilde{c}$ (see \S\ref{sec:iliev0}). We also explicitly verify this using the same test in \S\ref{sec:iliev0}, but with different $\tilde{c}$. Thus, the results in the main text are not affected by this issue.

\begin{figure}
\includegraphics[width=0.49\textwidth]{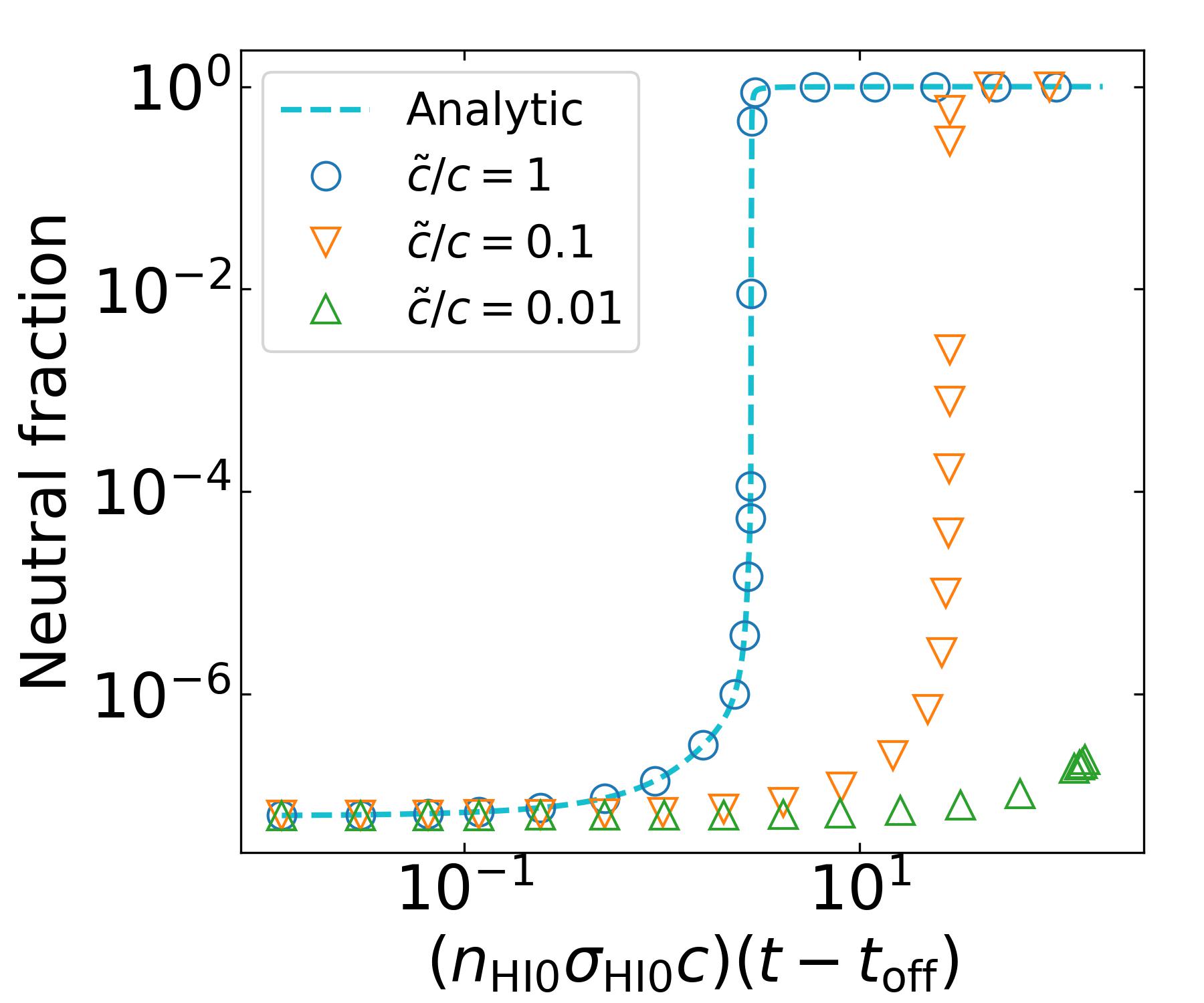}
\caption{Neutral fraction evolution in the test described in \S\ref{sec:iliev0} after the radiation source is turned off at $t_{\rm off}$. Unlike in the main text, we consider different reduced speeds of light $\tilde{c}$ in our simulations, as circles ($\tilde{c}=c$), inverted triangles ($\tilde{c}=0.1c$), and up pointing triangles ($\tilde{c}=0.01c$). Only the $\tilde{c}=c$ case (i.e., without the reduced speed of light approximation) can match the analytic solution. This demonstrates the limitation of the reduced speed of light approximation in this test. }
\label{fig:issueRSL}
\end{figure}

\section{Test of Equilibrium Calculations}
\label{sec:verequcal}

We verify the \chimes\ equilibrium solver, applied in \S\ref{sec:comequ}. The solver takes the element abundances, temperatures, densities, and radiation fields to calculate the equilibrium ion species abundances. 

We consider the ``\ion{H}{II} region with metal'' test as in \S\ref{sec:HIImetal} and obtain the equilibrium ion abundance from the \chimes\ equilibrium solver. As in \S\ref{sec:HIImetal}, we consider a time $t=1.28\;{\rm Myr}$, so the species fractions should be close to equilibrium in the inner region. 

In Fig.~\ref{fig:eqmpy}, we compare the nitrogen ion abundances from the simulated species fractions to those from the \chimes{} equilibrium solver. The differences between the \chimes{} equilibrium solver and the simulations are smaller: in general, less than 5\% in the inner region (<\;6\,pc) and at most 15\% near the ionization front (for \ion{N}{II}). This demonstrates our \chimes{} equilibrium solver can accurately estimate the equilibrium ion species.

\begin{figure}
\includegraphics[width=0.49\textwidth]{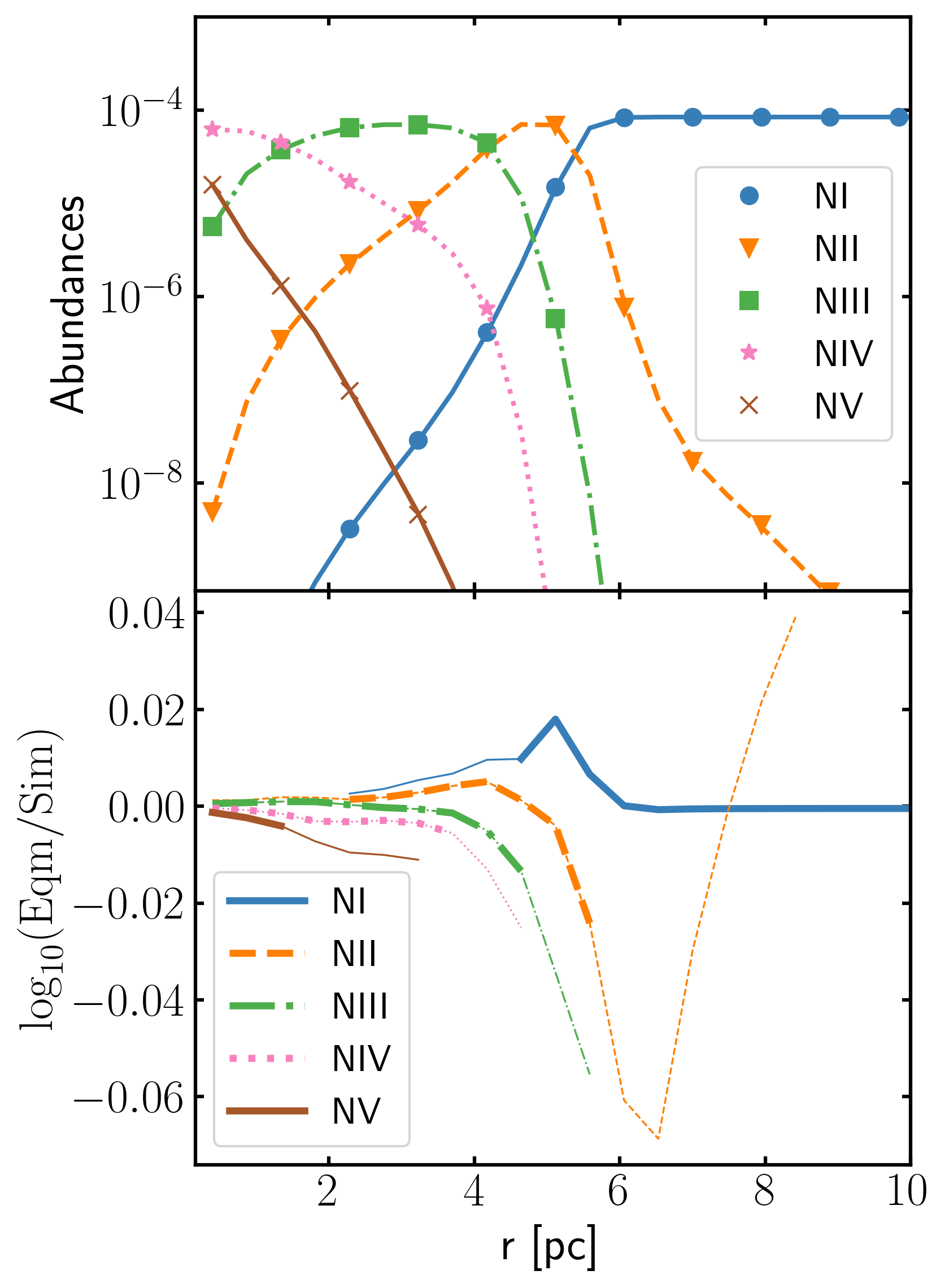}
\caption{Comparison of ion abundances between the non-equilibrium simulation and the \chimes\ equilibrium solver in the ``\ion{H}{II} region with metal'' test (as in \S\ref{sec:HIImetal}). {\it Upper panel:} Lines represent ion species abundances from the non-equilibrium simulation (\ion{N}{I}: blue solid; \ion{N}{II}: orange dashed; \ion{N}{III}: green dash-dot; \ion{N}{IV}: dotted; \ion{N}{V}: brown solid). Markers represent ion species abundances from the \chimes\ equilibrium solver. {\it Bottom panel:} Relative ratio of ion species abundances between the \chimes\ equilibrium solver (eqm) and non-equilibrium simulation (Sim). The thin lines show the ratio with (eqm) abundances $>10^{-9}$, whereas the thick lines show the ratio with (eqm) abundances $>10^{-6}$.}
\label{fig:eqmpy}
\end{figure}

\section{Test of multiple frequency bins in thermo-chemistry test in a single gas particle}
\label{sec:multibin_test}

\begin{figure}
\includegraphics[width=0.49\textwidth]{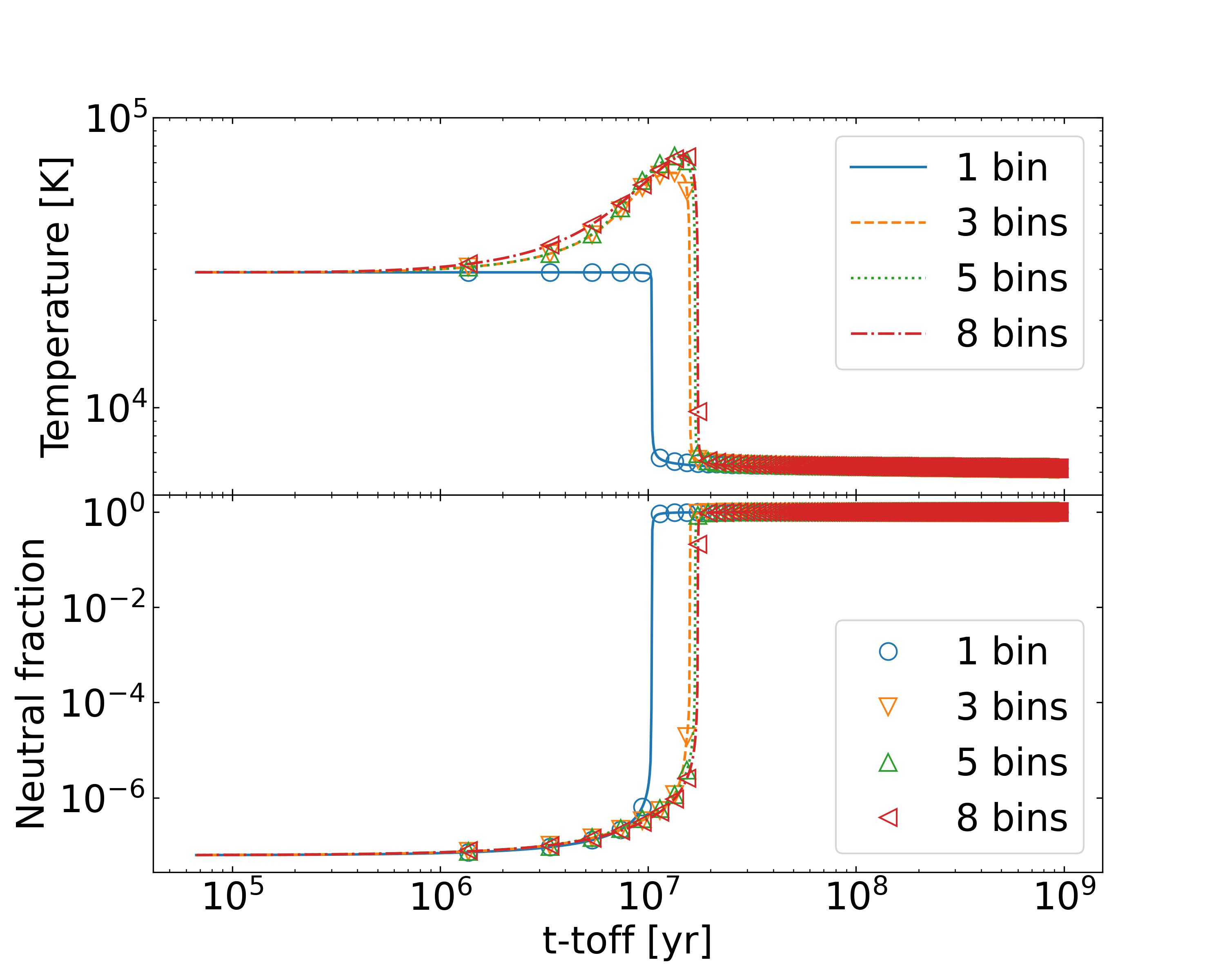}
\caption{Test 1: Gas and photon evolution after the radiation source is turned off (following \S\ref{sec:iliev0}).  The horizontal axis is the time measured from the source switched off, $t-t_{\rm off}$. The different colour symbols represent the results from \swifter{} using different numbers of spectral bins. The different lines represent the results from integrating the rate equations using the {\sc odeint} routine in {\sc scipy}. We find excellent agreements between \swifter{} and {\sc odeint}. The figure also shows that it is essential to include multiple spectral bins, since the ``1 bin'' solver underestimates the time to return to neutral and misses the rise of temperature after turning off the radiation.}
\label{fig:multibin_test}
\end{figure}

We examine the effects of different numbers of spectral bins on the gas evolution in Test 1 (\S\ref{sec:iliev0}; an optically thin gas particle is photo-heated by multi-frequency UV radiation and then the radiation is turn off.). We test the following spectral bins, with spectral bin edges:
\begin{itemize}[label={}]
    \item 1 bin : [13.6, $\infty$] eV
    \item 3 bins: [13.6, 24.6, 54.4, $\infty$] eV
    \item 5 bins: [13.6, 24.6, 35.5, 75.0, $\infty$] eV
    \item 8 bins: [13.6, 18.0, 24.6, 35.5, 54.4, 68.0, 81.6, 95.2,  $\infty$] eV.
\end{itemize}
The ``1 bin'' and ``3 bins'' cases are identical to \S\ref{sec:iliev0}, whereas the ``5 bins'' and ``8 bins'' cases were applied in \S\ref{sec:HII} and \S\ref{sec:HIImetal} respectively. 

Fig. \ref{fig:multibin_test} shows the temperature and neutral fraction evolution after the radiation source is turned off. As in \S\ref{sec:iliev0}, we find different behaviours between single bin and multi-bin solutions: the multi-bin solutions have an increase in temperature due to spectral hardening. The multi-bin solutions also take longer to return to neutral due to lower photo-ionisation cross sections of more energetic photons.

Finally, we find that, at least with hydrogen only, three frequency bins are adequate to capture these multi-frequency effects, accurate to around 10-20 \%.

\label{lastpage}
\bsp
\end{document}